\documentclass[aps,prb,reprint,superscriptaddress]{revtex4-2}
 
\usepackage[T1]{fontenc}
\usepackage[utf8]{inputenc}
\usepackage{amsmath,amssymb,amsfonts,mathtools,bm}
\usepackage{physics}
\usepackage{graphicx}
\usepackage{xcolor}
\usepackage{hyperref}
\usepackage{microtype}
\usepackage{booktabs}
\usepackage{array}
\usepackage{multirow}
\usepackage{bbm}
\usepackage{quantikz}
\usepackage{amsthm}
\usepackage[ruled]{algorithm2e}
\usepackage[bb=dsserif]{mathalpha}
 
\hypersetup{
    colorlinks=true,
    linkcolor=blue,
    citecolor=blue,
    urlcolor=blue
}
\usepackage[normalem]{ulem}

\begin{document}

\title{Exploring Variational Regularization of Krylov Subspaces}

\author{Alexandre Perrin}
\email{alexandre.perrin.1@u-bordeaux.fr}
\affiliation{Université de Bordeaux, CNRS, LOMA, UMR 5798, F-33400 Talence, France}
\author{Jérémie Messud}
\affiliation{TotalEnergies, 2 place Jean Millier, 92078 Paris La Défense Cedex, France}
\author{Sébastien Burdin}
\affiliation{Université de Bordeaux, CNRS, LOMA, UMR 5798, F-33400 Talence, France}
\author{Matthieu Saubanère}
\affiliation{Université de Bordeaux, CNRS, LOMA, UMR 5798, F-33400 Talence, France}

\begin{abstract}
Krylov subspace methods efficiently approximate extremal eigenstates and eigenvalues of many-body Hamiltonians, but non-orthogonal bases can suffer from ill-conditioning. This challenge is amplified in quantum computing, where direct state access and thus straightforward orthogonalization are unavailable. We investigate a variational Krylov-like framework that iteratively optimizes a parametrized sequence of states to mimic Lanczos-like couplings while penalizing linear dependencies. Analyzed against exact Krylov and Lanczos on a finite Anderson impurity model, ansatz expressivity strongly influences the performance: Hamiltonian-inspired ansätze preserve Krylov-like convergence while improving the conditioning, whereas shallow ansätze remain stable but miss low-energy directions.
\end{abstract}

\maketitle
\raggedbottom
\section{Introduction}

Krylov subspace methods play a central role in the numerical solution of problems~\cite{golubMatrixComputations1996,parlettSymmetricEigenvalueProblem1998,saadNumericalMethodsLarge2011,haydockComputationalAdvantagesRecursion1984}. Among them, the Lanczos algorithm provides an exceptionally efficient way to approximate extremal eigenvalues and eigenstates of large Hamiltonians by constructing an orthonormal sequence of vectors that progressively captures the most relevant spectral directions~\cite{golubMatrixComputations1996,parlettSymmetricEigenvalueProblem1998,saadNumericalMethodsLarge2011}. This strategy underlies a wide range of techniques, from exact diagonalization and recursion methods~\cite{haydockComputationalAdvantagesRecursion1984} to density matrix renormalization group calculations and modern quantum algorithms.
However, such approaches are limited in terms of memory on classical devices, and Krylov subspace ideas have therefore recently been adapted to hybrid quantum-classical frameworks. In particular, quantum subspace expansion~\cite{mccleanHybridQuantumclassicalHierarchy2017,collessComputationMolecularSpectra2018}, quantum filter diagonalization~\cite{parrishQuantumFilterDiagonalization2019,cohnQuantumFilterDiagonalization2021}, and quantum Krylov algorithms~\cite{stairMultireferenceQuantumKrylov2020,jamet2021krylovvariationalquantumalgorithm,cortesQuantumKrylovSubspace2022,mottaSubspaceMethodsElectronic2024} have received increasing attention. In these approaches, a set of quantum states spanning a reduced subspace is generated and used to construct projected Hamiltonian and overlap matrices, whose generalized eigenvalue problem approximates the spectrum of the original Hamiltonian. Such methods are part of the broader family of near-term variational and hybrid quantum algorithms~\cite{peruzzoVariationalEigenvalueSolver2014,mccleanTheoryVariationalHybrid2016,bhartiNoisyIntermediatescaleQuantum2022,tillyVariationalQuantumEigensolver2022,bauerQuantumAlgorithmsQuantum2020}. Their performance depends not only on the expressivity of the generated states, but also on the numerical stability of the projected subspace problem.

A fundamental difficulty of Krylov-inspired constructions is the conditioning of the generated basis. Indeed, repeated applications of the Hamiltonian act as a spectral filter: components associated with dominant eigenvalues are amplified relative to the others. This mechanism explains the rapid convergence of Krylov methods, but it also causes successive Krylov vectors to become increasingly parallel as the iteration number grows. As a result, the overlap matrix $S$ associated with the non-orthogonal basis constructed by applying successive powers of the Hamiltonian to an initial state,
$\left\{
\ket{\phi_0},
H\ket{\phi_0},
\ldots,
H^n\ket{\phi_0}
\right\}$
develops very small eigenvalues. A standard measure of the non-orthogonality is the condition number defined as the ratio between the largest and the smallest eigenvalues of $S$ that is shown to grow rapidly with the subspace dimension. The corresponding generalized eigenvalue problem then becomes increasingly sensitive to numerical errors~\cite{saadNumericalMethodsLarge2011,parlettSymmetricEigenvalueProblem1998,baiTemplatesSolutionAlgebraic2000}.

This conditioning issue is precisely why classical Krylov algorithms are not usually implemented in the non-orthogonal monomial basis. Instead, Lanczos and Arnoldi algorithms add an {\it on the fly} orthogonalization step in the Krylov algorithm to recursively construct an orthonormal basis, thereby avoiding the explicit use of the ill-conditioned Krylov vectors~\cite{saadNumericalMethodsLarge2011,parlettSymmetricEigenvalueProblem1998}. In finite precision arithmetic, the loss of orthogonality in Lanczos-type recursions is itself a central numerical issue~\cite{paigeErrorAnalysisLanczos1976}. Additional procedures such as partial, selective, or full reorthogonalization, deflation, thresholding of nearly linearly dependent directions, and restarted Krylov schemes are often required to maintain numerical stability and control the size of the generated subspace~\cite{parlettLanczosAlgorithmSelective1979,simonLanczosAlgorithmPartial1984,lehoucqARPACKUsersGuide1998,baiTemplatesSolutionAlgebraic2000}.

In quantum-computing settings, the same issue appears in a different form. 
Although the quantum state is physically prepared on the processor, it is not available as an explicit classical state vector. 
As a result, the operations used in a classical orthogonalization procedure (e.g. Gram-Schmidt), such as projecting a vector onto previously generated directions, subtracting these components, and renormalizing the residual state, cannot be straightforwardly adapted in quantum circuits without explicit memory (formal elements are given in Appendix \ref{app:GS}).
Instead, as of today, quantum subspace methods typically estimate overlap and Hamiltonian matrix elements, and the orthogonalization is performed only at the level of the measured projected matrices. 
This classical post-processing is sufficient in principle, but becomes numerically unstable when the prepared states are nearly linearly dependent: small eigenvalues of the overlap matrix amplify shot noise, finite-sampling errors, and other estimation errors in the generalized eigenvalue problem~\cite{hugginsNonorthogonalVariationalQuantum2020,parrishQuantumFilterDiagonalization2019,cohnQuantumFilterDiagonalization2021,kirbyAnalysisQuantumKrylov2024,zhangMeasurementefficientQuantumKrylov2024,nakamuraAdaptiveMeasurementStrategy2024,getelinaQuantumSubspaceExpansion2024a}. 
The interplay between the conditioning of the overlap matrix and the noise level has been analyzed in detail for quantum subspace diagonalization, where thresholding of the smallest overlap eigenvalues is shown to act as an effective regularization~\cite{epperlyTheoryQuantumSubspace2022,klymkoRealTimeEvolutionUltracompact2022}.
Recent work has also explored constructing quantum subspaces that are orthogonal by design, thereby avoiding overlap measurements and the associated conditioning problem~\cite{izmaylovClassicalActiveSpaceHybrid2026}.
This motivates the study of alternative constructions that generate useful non-orthogonal subspaces while seeking to control their conditioning during state preparation, rather than only regularizing the projected matrices after measurement.
Several subspace-based quantum algorithms have explored related ideas. 
Non-orthogonal variational quantum eigensolvers formulate the eigenvalue problem directly in a measured non-orthogonal basis~\cite{hugginsNonorthogonalVariationalQuantum2020}. 
Quantum filter diagonalization constructs projected subspaces from time-evolved or filtered states and solves the corresponding generalized eigenvalue problem~\cite{parrishQuantumFilterDiagonalization2019,cohnQuantumFilterDiagonalization2021}. 
Quantum Krylov methods use Hamiltonian-generated or time-evolved Krylov-like states to build low-dimensional projected problems~\cite{stairMultireferenceQuantumKrylov2020,kirbyExactEfficientLanczos2023,kirbyAnalysisQuantumKrylov2024,zhangMeasurementefficientQuantumKrylov2024,klymkoRealTimeEvolutionUltracompact2022,shenRealTimeKrylovTheory2023,stairStochasticQuantumKrylov2023,francisSubspaceDiagonalizationQuantum2022,yoshiokaKrylovDiagonalizationLarge2025}. 
Iterative quantum-assisted eigensolvers instead enlarge the variational space through operator-generated states selected during the algorithm~\cite{bhartiIterativeQuantumassistedEigensolver2021}. 
Variational subspace constructions generate the subspace through optimized parametrized states rather than through a fixed Krylov recursion~\cite{akandeVariationalQuantumSubspace2025,jamet2021krylovvariationalquantumalgorithm}. 
These works do not all attempt to implement explicit orthogonalization on a quantum processor; rather, they rely on measured subspace matrices, filtering procedures, operator-generated expansions, or variationally generated states to construct a useful low-dimensional projected problem. 
They illustrate the potential of subspace-based quantum algorithms, but they also highlight a general challenge: generating states that are both physically relevant and sufficiently linearly independent is nontrivial. 
Good convergence and good conditioning are not automatically compatible requirements.

In this work, we investigate a variational Krylov-like construction designed to probe this compromise. Instead of generating basis states by repeated applications of the Hamiltonian, we iteratively construct a sequence of parametrized states. Each new state is optimized according to a cost function inspired by the off-diagonal couplings of the Lanczos recursion, while a regularization term penalizes overlap with previously generated states. The resulting basis is not an exact Krylov basis, nor is it guaranteed to reproduce the Lanczos recursion. Rather, it defines a regularized Krylov-like subspace whose convergence and conditioning can be compared to exact Krylov and Lanczos references.
The objective of the present work is therefore not to establish a generally scalable eigensolver, but to perform a controlled numerical study of the trade-off between convergence, conditioning, ansatz structure, and optimization stability. We benchmark the construction on a finite Anderson impurity model in star geometry, where exact diagonalization and exact Lanczos references are available. This allows us to analyze how different variational ansätze reproduce, regularize, or fail to reproduce Krylov-generated directions.

The remainder of this paper is organized as follows. Section~\ref{sec:framework} recalls the Krylov and non-orthogonal subspace framework and the origin of the conditioning problem. Section~\ref{sec:direct_lanczos} discusses a direct coherent construction of the Lanczos recursion and its recursive cost. Section~\ref{sec:method} introduces the variational construction, the regularized cost function, and the families of ansätze considered. Section~\ref{sec:model} defines the Anderson impurity benchmark and the numerical protocol. Section~\ref{sec:results_discussion} presents the numerical analysis, and Sec.~\ref{sec:conclusion} summarizes our conclusions. The recursive orthonormalization formulas are collected in Appendix~\ref{app:orthogonalization}.

\section{Krylov Subspace Methods}
\label{sec:framework}


Krylov subspace methods constitute a central class of techniques for solving large-scale eigenvalue problems~\cite{golubMatrixComputations1996,parlettSymmetricEigenvalueProblem1998,saadNumericalMethodsLarge2011}. Given a Hamiltonian \(H\) and an initial state \(\ket{\phi_0}\), the associated Krylov subspace of dimension \(D\) is defined as
\begin{equation}
    \mathcal K_D(H,|\phi_0\rangle)
=
\mathrm{span}\left\{
|\phi_0\rangle,H|\phi_0\rangle,\ldots,H^{D-1}|\phi_0\rangle
\right\}.
\end{equation}
In practice, Krylov subspaces provide highly efficient approximations of extremal eigenstates, as the repeated action of \(H\) progressively amplifies the components of \(\ket{\phi_0}\) along the dominant eigenvectors.
The effectiveness of Krylov methods can be understood from a simple spectral argument.
Let
\[
H=\sum_i \lambda_i |v_i\rangle\langle v_i|,
\]
with eigenvalues ordered such that
\[
|\lambda_0| > |\lambda_1| \ge |\lambda_2| \ge \cdots.
\]
Repeated applications of the Hamiltonian generate
\[
H^n|\phi_0\rangle
=
\sum_i c_i \lambda_i^n |v_i\rangle , 
\]
where \(c_i=\langle v_i|\phi_0\rangle\).
As \(n\) increases, the relative weight of the dominant eigenvalue grows as
\[
\left|\frac{\lambda_i}{\lambda_0}\right|^n ,
\]
leading to an exponential suppression of subleading spectral components.
\begin{figure}[t]
    \centering
    \includegraphics[width=\columnwidth]{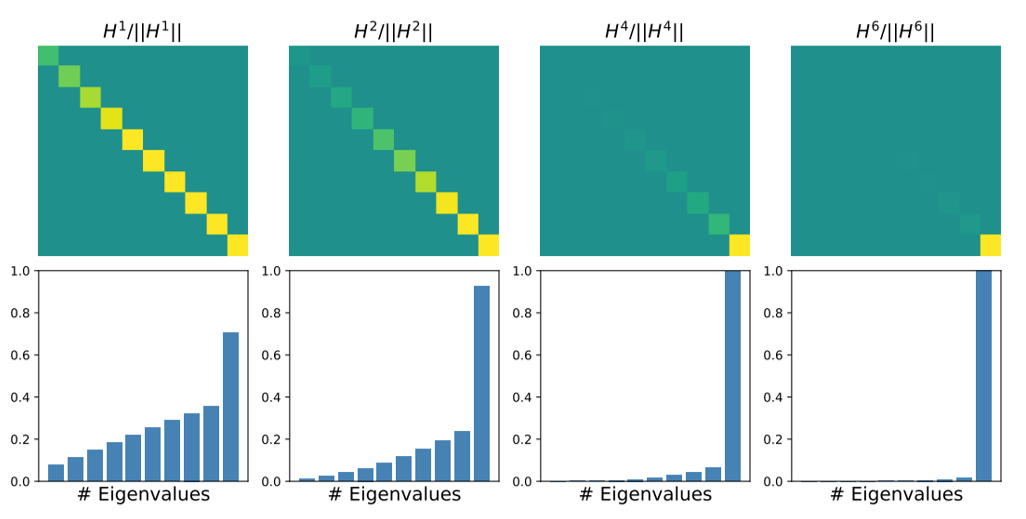}
    \caption{
Illustration of the spectral filtering mechanism underlying Krylov and power-method constructions.
Top row: normalized matrices (taken in a diagonal form with ascending values) $H^n/\|H^n\|$ for increasing powers $n$.
Bottom row: normalized eigenvalue spectra associated with these matrices.
Repeated applications of the Hamiltonian amplify the dominant spectral component while exponentially suppressing subleading eigenvalues.
This progressive concentration onto a small number of directions explains both the rapid convergence of Krylov-based methods and the deterioration of the conditioning of the corresponding subspaces.
}
    \label{fig:power}
\end{figure}
Figure~\ref{fig:power} illustrates this spectral filtering mechanism, known as the power method \cite{haydockComputationalAdvantagesRecursion1984,saadNumericalMethodsLarge2011,sekiQuantumPowerMethod2021}.
The upper panels show the normalized matrices (taken in a diagonal form with ascending values) \(H^n/\|H^n\|\) for increasing powers, while the lower panels display the corresponding normalized eigenvalue spectra.
As the power increases, the largest eigenvalue progressively dominates the spectrum and the matrix becomes increasingly rank-one-like.
This concentration is responsible for the efficiency of power and Krylov methods; it also implies that successive Krylov vectors become progressively more aligned.
The resulting loss of linear independence is the fundamental origin of the conditioning issues encountered in Krylov subspace constructions.


Subspace methods provide a natural extension to exploit Krylov-like constructions in practical calculations. Given the set of non-orthogonal Krylov vectors  \(\{\ket{\phi_i} = H^i \ket{\phi_0}\}\) where $\ket{\phi_0}$ is an arbitrary initial vector, one defines the projected Hamiltonian in the Krylov subspace and overlap matrix as
\begin{align}
H_{ij} &= \bra{\phi_i} H \ket{\phi_j},
&
S_{ij} &= \braket{\phi_i}{\phi_j} .
\label{eq:projected_matrices}
\end{align}
The same symbol is used for the full Hamiltonian $H$ acting on the many-body Hilbert space and for its projection onto the subspace, the distinction following from the presence of subspace indices; in matrix expressions such as Eq.~\eqref{eq:hperp} below, $H$ always denotes the projected matrix of Eq.~\eqref{eq:projected_matrices}. The approximate eigenvalue problem is then formulated as a generalized eigenvalue equation,
\begin{equation}
H \bm{c}
=
E  S  \bm{c},
\label{eq:gep}
\end{equation}
where $\bm{c}$ collects the expansion coefficients of the approximate eigenstate in the non-orthogonal basis. Its solutions provide approximations to the eigenstates of the full Hamiltonian within the chosen subspace.

This formulation encompasses a broad class of approaches, including quantum subspace expansion and related quantum equation-of-motion methods~\cite{mccleanHybridQuantumclassicalHierarchy2017,collessComputationMolecularSpectra2018,ollitraultQuantumEquationMotion2020,yoshiokaGeneralizedQuantumSubspace2022,getelinaQuantumSubspaceExpansion2024a}, where the states \(\{\ket{\phi_i}\}\) are generated from a reference state using a set of excitation operators. More generally, the accuracy of the method depends on the ability of the chosen subspace to capture the relevant physical directions in Hilbert space. A central aspect of this formulation is the role of the overlap matrix \(S\). In contrast to orthonormal bases, where \(S\) reduces to the identity, a non-orthogonal basis introduces additional structure that must be handled explicitly. In particular, the conditioning of \(S\) plays a crucial role in the stability of the generalized eigenvalue problem \cite{golubMatrixComputations1996,parlettSymmetricEigenvalueProblem1998,saadNumericalMethodsLarge2011}. If the basis vectors become nearly linearly dependent, \(S\) develops very small eigenvalues, and the inversion or factorization of \(S\) becomes numerically unstable.
To make this structure explicit, it is often convenient to transform the problem to an orthonormal basis. Using, for instance, a Cholesky decomposition of the overlap matrix,
\begin{equation}
    S = L L^\dagger,
    \label{eq:smatrix}
\end{equation}
one can define, in the case where $L$ is invertible, an effective Hamiltonian
\begin{equation}
    H^{\perp} = L^{-1} H L^{-\dagger},
    \label{eq:hperp}
\end{equation}
which is Hermitian and expressed in an orthonormal basis
and where we denote $L^{-\dagger}=(L^{\dagger})^{-1}=(L^{-1})^\dagger$.
The eigenvalues of \(H^{\perp}\) coincide with those of the generalized problem, Eq.~\eqref{eq:gep}, and the eigenvectors become $L^\dagger \bm{c}$. They are orthonormal by construction, their overlap matrix becoming the identity. 

This transformation is exact provided that $S$ is positive definite. However, when $S$ becomes nearly singular, the inverse Cholesky factors amplify numerical errors, while a rank-deficient $S$ makes the transformation undefined. Standard approaches address this issue through regularization or truncation of the overlap matrix, potentially modifying the effective subspace problem. Here, we instead investigate a complementary strategy: controlling the conditioning of $S$ directly during the variational construction of the basis, rather than relying on an a posteriori regularization of the projected matrices.

\begin{figure}[t]
    \centering
    \includegraphics[width=\columnwidth]{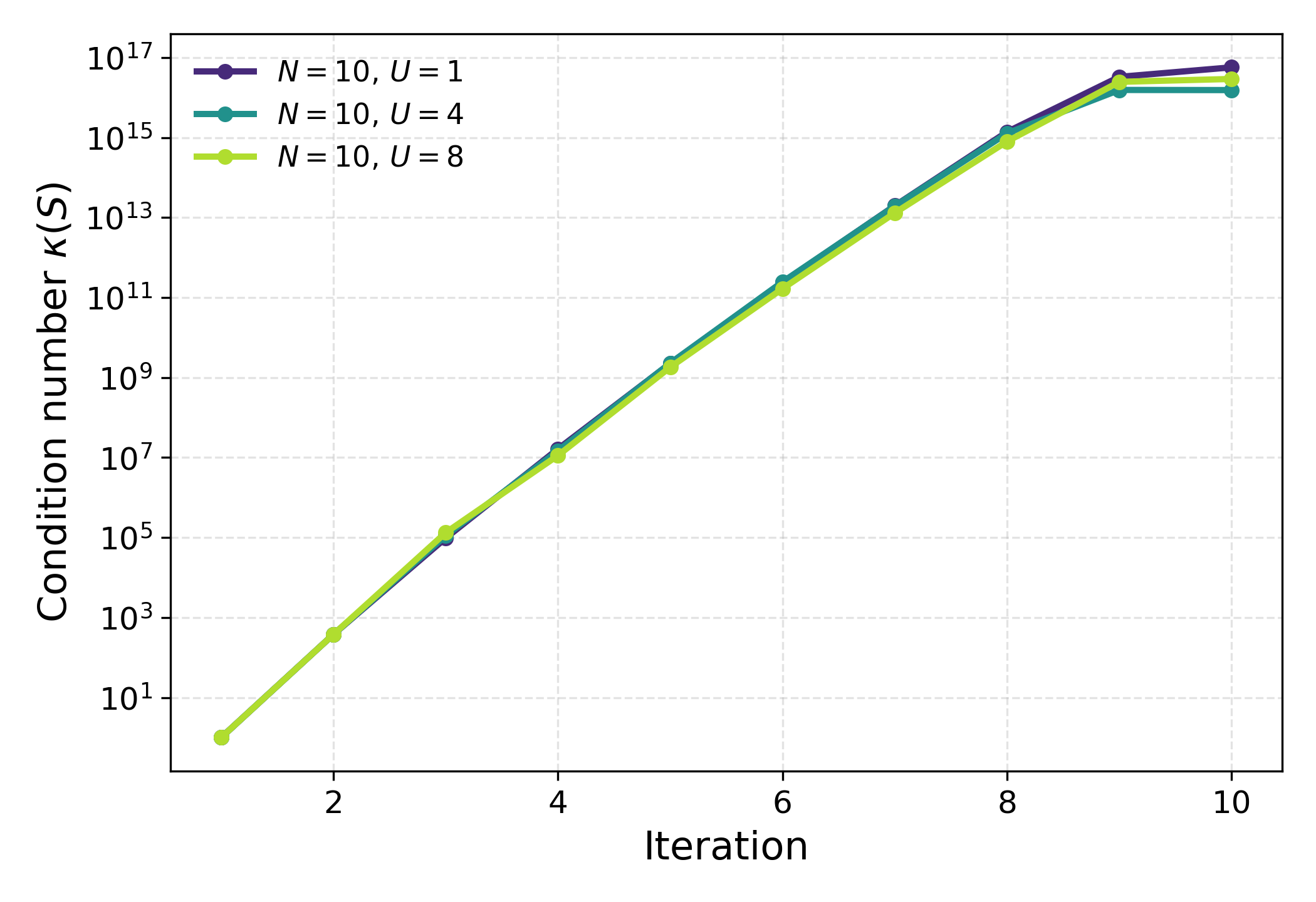}
    \caption{
Condition number of the overlap matrix associated with the Krylov basis
\(
\{|\phi_0\rangle, H|\phi_0\rangle, \dots, H^n|\phi_0\rangle\}
\)
as a function of the iteration number.
The exponential growth of \(\kappa(S)\) reflects the rapid emergence of
near-linear dependencies within the Krylov subspace, essentially independently
of the interaction strength.
Data shown for the \(N=10\) Anderson impurity model in star geometry \cite{PhysRev.124.41}
at \(U=1\), \(4\) and \(8\).
}
    \label{fig:krylov_condition}
\end{figure}

Within this perspective, the quality of a subspace construction is determined not only by its ability to approximate the target eigenstates, but also by the spectral properties of the overlap matrix \(S\). In particular, the smallest eigenvalue \(\lambda_{\min}(S)\) and the condition number
\begin{equation}
    \kappa(S) = \frac{\lambda_{\max}(S)}{\lambda_{\min}(S)}
    \label{eq:condition_number}
\end{equation}
provide quantitative measures of the linear independence of the basis and of the numerical stability of the projected problem. For a Krylov basis, $\kappa(S)$ increases rapidly, often exponentially, with the number of iterations, as shown in Fig.~\ref{fig:krylov_condition}.

This behavior has direct numerical consequences. A poorly conditioned overlap matrix leads to instabilities in the solution of the generalized eigenvalue problem, as well as to a loss of accuracy in the extracted eigenvalues and eigenvectors~\cite{epperlyTheoryQuantumSubspace2022,wilkinsonAlgebraicEigenvalueProblem1992}. In practice, this limits the usable dimension of Krylov subspaces, even when the underlying method would otherwise provide highly accurate approximations, and it is the reason why classical Krylov algorithms build orthonormal bases recursively rather than solving Eq.~\eqref{eq:gep} in the non-orthogonal basis.
From a broader perspective, this reveals a fundamental trade-off in subspace constructions: directions in Hilbert space that are most relevant for improving the variational energy are often those that are already partially explored by the existing basis, and therefore tend to introduce linear dependencies. Consequently, achieving both rapid convergence and good numerical conditioning is a nontrivial task.
These observations motivate the development of alternative constructions that preserve the favorable convergence properties of Krylov methods while explicitly controlling the conditioning of the generated basis. Before introducing the variational framework, we first examine what happens when the orthogonal Lanczos recursion is translated directly into coherent state preparation.

%
\section{Direct construction of an orthogonal Krylov subspace}
\label{sec:direct_lanczos}

The classical Lanczos algorithm avoids the ill-conditioned monomial Krylov basis by constructing an orthonormal sequence directly. Starting from a normalized state $\ket{q_0}$, the three-term recurrence reads, for $n\geq 1$,
\begin{equation}
    \beta_n \ket{q_{n+1}}
    =
    H\ket{q_n}
    -\alpha_n\ket{q_n}
    -\beta_{n-1}\ket{q_{n-1}},
    \label{eq:direct_lanczos_recursion}
\end{equation}
with
\begin{equation}
\begin{aligned}
    \alpha_n &= \mel{q_n}{H}{q_n},\\
    \beta_n &= \mel{q_{n+1}}{H}{q_n},
\end{aligned}
\label{eq:direct_lanczos_coefficients}
\end{equation}
The first step is given separately below. We now ask what happens if this recurrence is translated literally into coherent state preparation.  In the following, $H$ is kept as an abstract linear-operation primitive: we focus only on the recursive structure of the state preparation, independently of the particular realization of $H$ on a quantum processor.

\subsection{First Lanczos vector}

For the first step,
\begin{equation}
    \ket{q_1}
    =\frac{1}{\beta_0}(H-\alpha_0 I)\ket{q_0}.
    \label{eq:direct_q1}
\end{equation}
For clarity, let us assume $\alpha_0\geq0$; a negative coefficient only changes the relative phase of the corresponding branch. Defining
\begin{equation}
    S_0=1+\alpha_0,
    \qquad
    \theta_0=\arccos\sqrt{\frac{1}{S_0}},
\end{equation}
one ancilla is sufficient to form the required linear combination.  After preparing the two branches, changing their relative sign, applying $H$ on the $\ket{0}$ branch, and unpreparing the ancilla, the useful branch is
\begin{equation}
    (\bra{0}\otimes I)\ket{\Psi_1}
    =\frac{H-\alpha_0 I}{S_0}\ket{q_0}
    =\frac{\beta_0}{S_0}\ket{q_1}.
    \label{eq:direct_q1_good_branch}
\end{equation}
Here, $\otimes$ denotes the tensor product between the ancilla and system registers.
The corresponding schematic quantum circuit is
\begin{equation}
\begin{quantikz}[column sep=0.25cm]
\lstick{$\ket{0}$}
& \gate{R_y(2\theta_0)}
& \gate{Z}
& \gate{X}
& \ctrl{1}
& \gate{X}
& \gate{R_y(-2\theta_0)}
& \qw \\
\lstick{$\ket{q_0}$}
& \qw
& \qw
& \qw
& \gate{H}
& \qw
& \qw
& \qw
\end{quantikz}
\label{eq:direct_q1_circuit}
\end{equation}
Thus, already the first Lanczos step requires a coherent linear combination and a selection of the desired ancilla branch.

\subsection{Opening the second step}

The second Lanczos vector is
\begin{equation}
    \ket{q_2}
    =\frac{1}{\beta_1}
    \left(
    H\ket{q_1}-\alpha_1\ket{q_1}-\beta_0\ket{q_0}
    \right).
    \label{eq:direct_q2}
\end{equation}
At a compact level, this requires three branches proportional to $H\ket{q_1}$, $\ket{q_1}$, and $\ket{q_0}$. The important point becomes apparent when the preparation of $\ket{q_1}$ is opened explicitly.  Two copies of the first-step circuit are required, one for $H\ket{q_1}$ and one for $\ket{q_1}$, in addition to the direct preparation of $\ket{q_0}$.  Consequently, the fully unfolded second step already contains three occurrences of the initial-state preparation.

To display only this recursive structure, we denote by $R_0^{\pm}$ the preparation/unpreparation rotations of Eq.~\eqref{eq:direct_q1_circuit}, while $R_1^{\pm}$ and $T_1^{\pm}$ denote the outer selector rotations. Their angles are chosen so that the success branch reproduces the relative coefficients in Eq.~\eqref{eq:direct_q2}, including the normalization of the embedded $q_1$ preparation. A fully unfolded schematic quantum circuit is then
\begin{figure}[ht]
\centering
\resizebox{0.98\linewidth}{!}{%
\begin{quantikz}[column sep=0.11cm,row sep=0.16cm]
\lstick{$\ket{0}$}
& \qw
& \gate{T_1^+}
& \qw
& \octrl{3}
& \octrl{2}
& \octrl{2}
& \octrl{3}
& \octrl{2}
& \octrl{3}
& \octrl{3}
& \octrl{2}
& \octrl{2}
& \octrl{3}
& \octrl{2}
& \ctrl{3}
& \gate{T_1^-}
& \qw
& \qw \\
\lstick{$\ket{0}$}
& \gate{R_1^+}
& \ctrl{-1}
& \gate{Z}
& \octrl{2}
& \octrl{1}
& \octrl{1}
& \octrl{2}
& \octrl{1}
& \octrl{2}
& \ctrl{2}
& \ctrl{1}
& \ctrl{1}
& \ctrl{2}
& \ctrl{1}
& \ctrl{2}
& \ctrl{-1}
& \gate{R_1^-}
& \qw \\
\lstick{$\ket{0}$}
& \qw
& \qw
& \qw
& \qw
& \gate{R_0^+}
& \gate{Z}
& \octrl{1}
& \gate{R_0^-}
& \qw
& \qw
& \gate{R_0^+}
& \gate{Z}
& \octrl{1}
& \gate{R_0^-}
& \qw
& \qw
& \qw
& \qw \\
\lstick{$\ket{0}$}
& \qw
& \qw
& \qw
& \gate{\mathcal Q_0}
& \qw
& \qw
& \gate{H}
& \qw
& \gate{H}
& \gate{\mathcal Q_0}
& \qw
& \qw
& \gate{H}
& \qw
& \gate{\mathcal Q_0}
& \qw
& \qw
& \qw
\end{quantikz}%
}
\caption{Schematic unfolding of the direct construction of $\ket{q_2}$ down to the initial-state preparation $\mathcal Q_0$.  All ancillas start in $\ket{0}$.  The two copies of the $q_1$ construction and the direct $q_0$ branch lead to three occurrences of $\mathcal Q_0$. The circuit is intended to expose the recursive structure; implementation-dependent details associated with realizing $H$ are left implicit.}
\label{fig:direct_q2_unfolded}
\end{figure}
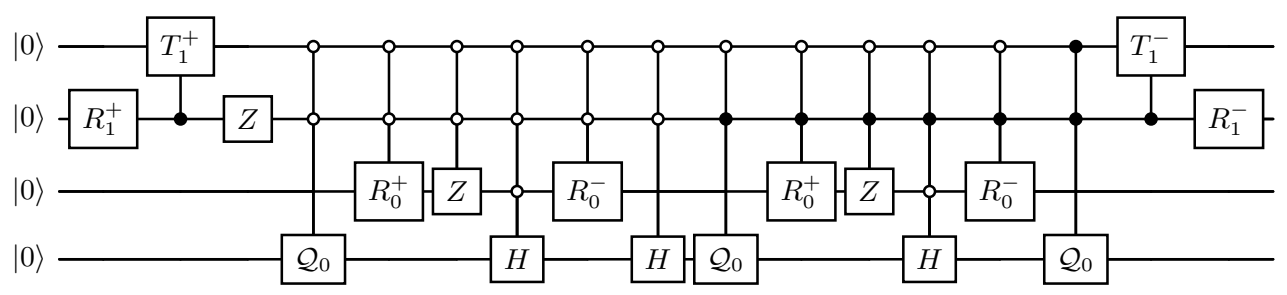

The purpose of Fig.~\ref{fig:direct_q2_unfolded} is not to propose an optimized compilation, but to make explicit what is hidden by the notation $\ket{q_1}$ in Eq.~\eqref{eq:direct_q2}: the previous state-preparation circuit must itself be coherently embedded in the next Lanczos step.

\subsection{General recursive step and cost}

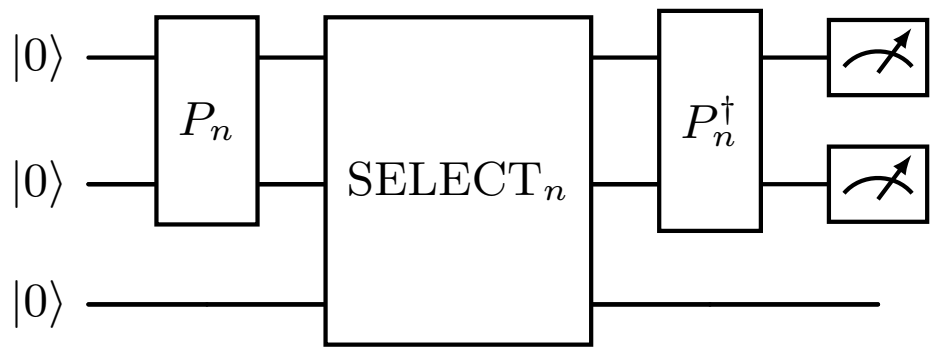
\begin{figure}[ht]
\centering
\begin{quantikz}[column sep=0.45cm,row sep=0.22cm]
\lstick{$\ket{0}$}
& \gate[2]{P_n}
& \gate[3]{\mathrm{SELECT}_n}
& \gate[2]{P_n^\dagger}
& \meter{} \\
\lstick{$\ket{0}$}
& \qw
& \qw
& \qw
& \meter{} \\
\lstick{$\ket{0}$}
& \qw
& \qw
& \qw
& \qw
\end{quantikz}
\caption{Folded representation of one direct Lanczos step.  The ancilla preparation $P_n$ encodes the three coefficients of the recurrence, while $\mathrm{SELECT}_n$ coherently selects the branches $H\mathcal Q_n$, $\mathcal Q_n$, and $\mathcal Q_{n-1}$.  Unpreparing the ancillas and selecting their $\ket{00}$ outcome isolates the component proportional to $\ket{q_{n+1}}$.}
\label{fig:direct_qn_folded}
\end{figure}

For a generic iteration it is convenient to keep the preparation folded.  Let $\mathcal Q_n$ denote a normalized preparation of $\ket{q_n}$. The direct translation of Eq.~\eqref{eq:direct_lanczos_recursion} has the schematic form
\begin{equation}
    \mathcal Q_{n+1}
    \sim
    \left[
    H\mathcal Q_n,\;
    \mathcal Q_n,\;
    \mathcal Q_{n-1}
    \right],
    \label{eq:direct_recursive_preparator}
\end{equation}
where an ancilla PREPARE/SELECT/UNPREPARE construction supplies the coefficients $1$, $-\alpha_n$, and $-\beta_{n-1}$ and projects onto the desired linear combination.  A compact representation is shown in Fig.~\ref{fig:direct_qn_folded}.  The two ancillas encode the three branches of the recurrence, while the block $\mathrm{SELECT}_n$ contains the controlled operations $H\mathcal Q_n$, $\mathcal Q_n$, and $\mathcal Q_{n-1}$.  In this literal construction, $\mathcal Q_n$ appears twice and $\mathcal Q_{n-1}$ once.

Formally, the selection block has the structure
\begin{equation}
\begin{aligned}
\mathrm{SELECT}_n \sim{}&
\ket{00}\!\bra{00}\otimes H\mathcal Q_n
+\ket{01}\!\bra{01}\otimes \mathcal Q_n\\
&+\ket{10}\!\bra{10}\otimes \mathcal Q_{n-1},
\end{aligned}
\label{eq:direct_select_n}
\end{equation}
with the relative signs and magnitudes carried by the ancilla amplitudes and phases prepared by $P_n$.  The fourth ancilla state is unused.

Let $N_n$ be the number of occurrences of the initial-state preparation $\mathcal Q_0$ after recursively opening all boxes.  From Eq.~\eqref{eq:direct_recursive_preparator},
\begin{equation}
    N_{n+1}=2N_n+N_{n-1},
    \qquad
    N_0=N_1=1.
    \label{eq:direct_cost_recursion}
\end{equation}
The first values are
\begin{equation}
    1,\;1,\;3,\;7,\;17,\;41,\ldots,
\end{equation}
and therefore
\begin{equation}
    N_n
    =\frac{(1+\sqrt2)^n+(1-\sqrt2)^n}{2}
    =\Theta\!\left[(1+\sqrt2)^n\right].
    \label{eq:direct_cost_scaling}
\end{equation}
The same recursive proliferation occurs for the applications of $H$.  Additional controls become more deeply nested as the recursion is opened, but their number along a single branch grows only linearly with $n$; the exponential growth above originates from the duplication of the previous state-preparation circuits themselves.

The precise exponential base depends on how aggressively the circuit is factorized. In particular, the two branches containing $\mathcal Q_n$ can be grouped by first preparing $\ket{q_n}$ once and then forming $(H-\alpha_n I)\ket{q_n}$. At the level of state-preparation calls this gives the reduced recursion $N_{n+1}=N_n+N_{n-1}$, and hence $N_n=\Theta(\varphi^n)$ with $\varphi=(1+\sqrt5)/2$. Thus, factorization improves the prefactor and the exponential base, but does not remove the exponential growth associated with recursively rebuilding the previous Lanczos vectors.

This estimate deliberately ignores the cost of obtaining the desired ancilla branch.  Postselection or amplitude amplification introduces an additional normalization-dependent overhead, so Eq.~\eqref{eq:direct_cost_scaling} should be viewed as the structural cost of the literal recursive realization rather than as a complete gate complexity estimate.

Importantly, this observation is not a no-go statement for quantum Lanczos methods.  The Lanczos vector can also be written as $\ket{q_n}\approx P_n(H)\ket{q_0}$, and a globally compiled polynomial transformation need not reproduce the recursive circuit tree shown above~\cite{rammalOrthogonalQuantumKrylov2026}.  The result instead shows that a direct coherent translation of the classical three-term recurrence is an unattractive route: the operation that is trivial classically---forming a new vector from two stored previous vectors---requires repeated coherent state preparation when the vectors exist only as quantum states.

There is also no obstruction at the level of existence. Since $\ket{q_n}$ and $\ket{q_{n+1}}$ are normalized, there always exists a (non-unique) unitary $V_n$ such that
\begin{equation}
    V_n\ket{q_n}=\ket{q_{n+1}}.
    \label{eq:exact_lanczos_unitary}
\end{equation}
The difficulty is to obtain a useful circuit representation of $V_n$ without first solving the state-preparation problem itself. This observation provides the link to the variational construction considered below: rather than implementing the exact recursive map, we approximate such a transformation by a parametrized circuit and optimize it to reproduce the relevant Lanczos-like couplings while controlling the linear dependence of the generated basis.

\section{Variational control of the non-orthogonality of Krylov subspace}
\label{sec:method}

\subsection{Iterative variational construction of the basis}
\label{subsec:iterative_construction}

We investigate a variational framework aimed at constructing Krylov-like subspaces while explicitly monitoring the conditioning of the resulting non-orthogonal basis. Rather than generating the subspace through repeated applications of the Hamiltonian, as in standard Krylov methods, we construct it iteratively using a parametrized ansatz.

Starting from an initial state $|\phi_0\rangle$, we define a sequence of states
\begin{equation}
    \ket{\phi_n(\bm{\theta}_n)}
    \propto
    \mathcal{A}(\bm{\theta}_n)\ket{\phi_{n-1}},
\end{equation}
where $\mathcal{A}(\bm{\theta}_n)$ denotes a parametrized transformation acting on the previously generated state. This sequential construction was motivated by preliminary tests in which each basis state was independently generated by applying a parametrized ansatz to the same initial state. Such an approach quickly became limited by the expressivity of the fixed ansatz. Applying each new transformation to the previously optimized state instead allows the accessible variational manifold to grow with the iterations. Each new state is obtained by optimizing the parameters $\bm{\theta}_n$ according to a cost function designed to favor the growth of a useful variational subspace.
This procedure differs from conventional variational eigensolvers, where a single parametrized state is optimized to approximate the target eigenstate. Here, the objective is instead to construct a \emph{sequence} of states whose span defines a reduced subspace for the projected eigenvalue problem. At iteration $n$ and after optimization, the working basis is therefore
\begin{equation}
\left(
\ket{\phi_0},
\ket{\phi_1(\bm{\theta}_1^{*})},
\dots,
\ket{\phi_n(\bm{\theta}_n^{*})}
\right).
\end{equation}
The newly generated state is then added to the existing basis, and the projected Hamiltonian and overlap matrices are updated accordingly.
This iterative construction is related in spirit to variational quantum subspace methods~\cite{nakanishiSubspacesearchVariationalQuantum2019,akandeVariationalQuantumSubspace2025}, non-orthogonal VQE~\cite{hugginsNonorthogonalVariationalQuantum2020}, and adaptive variational constructions~\cite{grimsleyAdaptiveVariationalAlgorithm2019,tangQubitADAPTVQEAdaptiveAlgorithm2021}.

The key ingredient of the construction is the cost function used to generate each new basis state. Its design is guided by two competing objectives: first, to retain part of the rapid convergence associated with Krylov and Lanczos methods; second, to avoid generating nearly linearly dependent states that would make the projected problem ill-conditioned.

In the following, we refer to the projected diagonalization in the variationally generated Krylov-like basis as variational quantum Krylov diagonalization (VQKD). The projected matrices $H^{VQKD}$ and $S$ are those of Eq.~\eqref{eq:projected_matrices}, evaluated on the variational states $\{\ket{\phi_n(\bm{\theta}_n)}\}$, and the associated orthonormal representation $H^{\perp}=L^{-1} H^{VQKD} L^{-\dagger}$ is defined as in Eqs.~\eqref{eq:smatrix} and~\eqref{eq:hperp}. 
Working with $H^\perp$ rather than $H$ allows us to characterize Hamiltonian couplings in an orthonormal basis, avoiding contributions arising solely from overlaps between nonorthogonal states.
All variational states are normalized, $\braket{\phi_n(\bm{\theta}_n)}{\phi_n(\bm{\theta}_n)}=1$, so that $S_{nn}=1$ at every iteration; this is enforced explicitly after applying the ansatz, which is required for the non-unitary constructions introduced in Sec.~\ref{subsec:ansatz_constructions}.

In this orthonormal representation, the off-diagonal element $(H^\perp)_{n,n-1}$ measures the effective Hamiltonian coupling between the newly added direction and the previous one. It is not, in general, an exact Lanczos coefficient, because the variational states are not constrained to satisfy a three-term Krylov recurrence. Nevertheless, it provides a natural proxy for a Lanczos-like coupling within the generated subspace.

At iteration $n$, the new state $|\phi_n^*\rangle=|\phi_n(\bm{\theta}_n^*)\rangle$ is obtained by minimizing the cost function
\begin{equation}
\mathcal{C}_n(\bm{\theta}_n)
=
-\left|
H^{\perp}_{n,n-1}(\bm{\theta}_n)
\right|^2
\left(
1
-
\frac{\lambda}{n}
\sum_{i<n}
\left|
\langle \phi_n(\bm{\theta}_n) | \phi_i^{*}\rangle
\right|^2
\right),
\label{eq:variational_lanczos_cost}
\end{equation}
where $\lambda\geq 0$ controls the strength of the regularization and $H^{\perp}_{n,n-1}(\bm{\theta}_n)$ denotes the $(n,n-1)$ element of $L^{-1}H^{VQKD}L^{-\dagger}$ evaluated for the trial state $\ket{\phi_n(\bm{\theta}_n)}$.
The first factor in Eq.~\eqref{eq:variational_lanczos_cost}, namely $-|H^{\perp}_{n,n-1}(\bm{\theta}_n)|^2$, favors states that maximize the effective Hamiltonian coupling with the previous orthonormalized basis direction. This term is the Lanczos-inspired part of the criterion: it encourages the ansatz to generate directions that are strongly connected to the current subspace by the Hamiltonian~\cite{akandeVariationalQuantumSubspace2025}.
The second, multiplicative factor,
\begin{equation}
\label{eq:2factor}
1
-
\frac{\lambda}{n}
\sum_{i<n}
\left|
\langle \phi_n (\bm{\theta}_n) | \phi_i^{*}\rangle
\right|^2 ,
\end{equation}
penalizes overlap between the newly generated state and the previously constructed basis states. Since the cost function is minimized, a large overlap reduces the gain associated with a large Hamiltonian coupling. This factor therefore acts as a soft constraint against near-linear dependencies and can improve the conditioning of the overlap matrix, although small pairwise overlaps alone do not guarantee good conditioning. The normalization by $n$ keeps the typical scale of the penalty approximately stable as the basis size increases. The multiplicative form was preferred over an additive regularization term based on preliminary numerical tests. Indeed, the magnitude of the effective Hamiltonian coupling varies across iterations, so that an additive penalty with fixed $\lambda$ does not maintain a consistent balance between the two contributions and may require retuning. The multiplicative form instead scales the penalty with the Hamiltonian coupling, allowing a more consistent weak regularization throughout the construction.

The parameter $\lambda$ controls the strength of the penalty. Since $\sum_{i<n}|\braket{\phi_n}{\phi_i}|^{2}\leq n$ for normalized states, Eq.~(\ref{eq:2factor}) remains nonnegative for $0\leq\lambda\leq 1$. For $\lambda>1$, it can change sign, losing its interpretation as an attenuation of the Hamiltonian-coupling term. All calculations reported below therefore use $\lambda\leq 1$; the influence of this parameter is studied below.

The two tendencies encoded in the cost function are inherently competing. The Hamiltonian-coupling term alone does not distinguish between a genuinely new direction and a direction whose dominant component is already contained in the existing subspace. Such a state may have a large effective coupling while contributing little new independent information, thereby worsening the conditioning of $S$. Conversely, a state that is nearly orthogonal to all previous states is not necessarily physically relevant. In particular, if the regularization is too strong, the optimization may favor well-conditioned but weakly useful directions. The parameter $\lambda$ therefore controls a compromise between following Krylov-like Hamiltonian-connected directions and maintaining a numerically stable basis.

As a result, the proposed scheme should be viewed as a \emph{variationally regularized Krylov-like construction}. It does not enforce the exact Lanczos recursion at the level of the generated states. Instead, it uses a Lanczos-inspired criterion within a constrained variational manifold, while explicitly penalizing the near-linear dependencies responsible for the ill-conditioning of  Krylov bases.

Finally, the use of $H^\perp$ provides a convenient way to interpret the non-orthogonal variational basis in terms of an associated orthonormal representation. This transformation is equivalent to orthonormalizing the projected basis at the matrix level, rather than explicitly performing a Gram--Schmidt procedure on the full Hilbert-space states. Such a formulation is particularly natural in quantum subspace methods, where the states themselves are not directly manipulated classically and the projected matrices are reconstructed from measured overlaps and Hamiltonian matrix elements~\cite{mccleanHybridQuantumclassicalHierarchy2017,hugginsNonorthogonalVariationalQuantum2020,mottaSubspaceMethodsElectronic2024}. The recursive formulas used in this work to update this orthonormal representation are provided in Appendix~\ref{app:orthogonalization}.

\subsection{Ansatz constructions considered}
\label{subsec:ansatz_constructions}

The behavior of the variational Krylov construction depends strongly on the parametrized transformation $\mathcal{A}(\bm{\theta})$ used to generate new basis states. In this work, we consider three families of ansätze with different levels of structure and variational freedom, summarized in Table~\ref{tab:ansatze}. These ansätze should not all be interpreted in the same way. The hardware-efficient and Trotterized ansätze correspond to standard unitary circuit constructions, while the Hamiltonian-inspired ansatz is used primarily as a diagnostic construction to assess how well a variational manifold aligned with the operator content of $H$ can reproduce Krylov-like directions.

This distinction is important from the point of view of quantum implementation. The Hamiltonian-inspired transformation considered below is not a generic shallow unitary circuit. Its direct implementation would require nontrivial linear-combination-of-unitaries or block-encoding primitives, and would therefore be substantially more demanding than standard NISQ-oriented ansätze such as hardware-efficient circuits or Trotterized variational Hamiltonian ansätze. In the present work, this construction is therefore mainly used to analyze the role of Hamiltonian structure, rather than to propose an immediately practical quantum circuit implementation.

\paragraph*{Hardware-efficient ansatz (HEA).}
As a problem-agnostic baseline, we first consider a hardware-efficient ansatz with no explicit relation to the Hamiltonian structure. Such ansätze are widely used in near-term variational algorithms, but their performance and trainability can depend strongly on the circuit architecture, the number of layers, and the system size~\cite{kandalaHardwareefficientVariationalQuantum2017,choyMolecularEnergyLandscapes2023,wangEHAEntanglementvariationalHardwareefficient2024}. In particular, increasing the number of layers can improve expressivity, but it also increases the number of variational parameters and may make the classical optimization substantially more difficult.

\textit{For the HEA baseline,} we use a real HEA composed of two layers of parametrized $R_y$ rotations separated by a single layer of nearest-neighbor CNOT gates arranged in a one-dimensional cascade. The ansatz can be written schematically as
\begin{equation}
\mathcal{A}_{\mathrm{HEA}}(\bm{\theta})
=
U_y(\bm{\theta_1}) U_{\mathrm{CNOT}}
U_y(\bm{\theta_2}),
\end{equation}
where $U_y(\bm{\theta})$ denotes a layer of single-qubit $R_y$ rotations and $U_{\mathrm{CNOT}}$ denotes the fixed CNOT cascade. The circuit therefore contains $2N_q$ variational parameters for $N_q$ qubits, and a single entangling layer, which keeps the optimization problem small and stable. This choice follows the same practical motivation as in Refs.~\cite{akandeVariationalQuantumSubspace2025,leonePracticalUsefulnessHardware2024}, where shallow HEA circuits are used to reduce circuit depth and mitigate optimization difficulties.

This construction provides a useful baseline because it can generate linearly independent states without explicitly encoding the Hamiltonian structure. However, this also means that good conditioning does not necessarily imply that the generated subspace is physically relevant for the low-energy sector. In the present study, the HEA is therefore used to separate the effect of conditioning from the effect of Hamiltonian-informed state generation.

\paragraph*{Variational Hamiltonian ansatz (VHA).}
We next consider a variational Hamiltonian ansatz based on Trotterized Hamiltonian evolution. The VHA is a problem-inspired unitary ansatz and has been used extensively for Hubbard-type systems~\cite{cadeStrategiesSolvingFermiHubbard2020,martinSimulatingStronglyInteracting2022,alvertisClassicalBenchmarksVariational2024}. In this construction, the Hamiltonian is decomposed into a sum of terms or groups of terms,
\begin{equation}
H = \sum_a H_a ,
\end{equation}
and the variational transformation is written as a product of exponentials,
\begin{equation}
\mathcal{A}_{\mathrm{VHA}}(\bm{\theta})
=
\prod_a
\exp\left(-i \tau \theta_a H_a\right).
\end{equation}
In all VHA calculations presented here, we use $\tau=0.1$. The parameter $\tau$ fixes the typical scale of the elementary Hamiltonian-generated rotations. Strictly speaking, if all parameters $\theta_a$ were fully unconstrained and globally optimized, $\tau$ could be absorbed into a redefinition of the variational parameters, $\tilde{\theta}_a=\tau\theta_a$. In practice, however, with a fixed initialization, finite optimization budget, and local nonlinear optimization, $\tau$ controls the effective step size explored by each VHA layer.

For small $\tau$, each exponential remains close to the identity,
\begin{equation}
\exp\left(-i\tau \theta_a H_a\right)
=
1
-
i\tau \theta_a H_a
+
\mathcal{O}(\tau^2),
\end{equation}
so that one VHA layer generates states that are perturbatively close to the previous state. This regime is naturally connected to Krylov-like constructions, since the leading correction is linear in Hamiltonian terms. It can also improve the stability of the iterative procedure, but may slow down the exploration of the variational subspace because successive states remain close to each other.

Larger values of $\tau$ have the opposite effect. They allow a single VHA layer to move further away from the previous state and may therefore improve the exploration of Hilbert space with a small number of iterations. However, the generated state is then no longer dominated by the linear term in the expansion above: higher-order powers and nested products of Hamiltonian components become important. As a result, the connection with a local Krylov-like step becomes less direct. In addition, larger effective rotation angles can make the optimization landscape more oscillatory and more sensitive to initialization. In this work, we therefore use the relatively small value $\tau=0.1$ as a conservative choice that favors stable, local Hamiltonian-generated updates.

\paragraph*{Hamiltonian-inspired ansatz $H(\theta)$.}
We then consider a Hamiltonian-inspired ansatz designed to align the variational manifold more directly with the operator content of the Hamiltonian. Hamiltonian-inspired and adaptive ansätze are motivated by the idea that the terms appearing in the Hamiltonian provide a compact and physically meaningful set of directions in Hilbert space~\cite{grimsleyAdaptiveVariationalAlgorithm2019,tangQubitADAPTVQEAdaptiveAlgorithm2021,anastasiouTETRISADAPTVQEAdaptiveAlgorithm2024,akandeVariationalQuantumSubspace2025}.

In this construction, each iteration introduces a new parametrized transformation built from Hamiltonian terms. Schematically, one may write
\begin{equation}
\mathcal{A}_{H}(\bm{\theta})
=
\sum_a \theta_a P_a ,
\end{equation}
where the operators $P_a$ denote the elementary terms or Pauli strings entering the qubit representation of the Hamiltonian. Unlike the other ansätze considered here, $H(\boldsymbol{\theta})$ is generally non-unitary and would therefore require a probabilistic implementation on a quantum processor, for instance through an LCU-based block encoding, with an associated post-selection or amplitude-amplification overhead. At iteration $n$, the state is written as
\begin{equation}
\ket{\phi_n(\bm{\theta}_n)}
\propto
\mathcal{A}_{H}(\bm{\theta}_n)
\ket{\phi^*_{n-1}},
\end{equation}
where only the most recent parameter vector $\bm{\theta}_n$ is optimized, while the previously optimized state $\ket{\phi^*_{n-1}}$ is kept fixed. This ansatz defines a local update rule in the iterative construction, analogous in spirit to a single-step Krylov recursion.

Because the transformation is constructed from Hamiltonian terms, this ansatz is expected to generate directions that are strongly connected to the previous state by $H$. It is therefore well suited for testing whether the Lanczos-inspired cost function can reproduce Krylov-like behavior. At the same time, its direct implementation as a quantum circuit is nontrivial, since the transformation is generally not a simple shallow unitary, and the resulting state must be renormalized explicitly.

An important property of this construction is that the exact Krylov step belongs to its variational manifold. Writing the qubit representation of the Hamiltonian as $H=\sum_a h_a P_a$, the choice $\bm{\theta}=\bm{h}$ gives $\mathcal{A}_{H}(\bm{h})=H$, so that $\ket{\phi_n}\propto H\ket{\phi_{n-1}}$. Consequently, at vanishing regularization the minimization of $\mathcal{C}_n$ can recover the exact Krylov recursion, and the $H(\theta)$ construction at $\lambda=0$ reduces, up to normalization, to the Krylov basis. The exact Krylov reference used in Sec.~\ref{sec:results_discussion} therefore plays the role of the $\lambda=0$ limit of this ansatz, and the deviations observed at finite $\lambda$ measure the effect of the overlap penalty alone. This property does not hold for the unitary $VHA(\theta)$ and $HEA(\theta)$ constructions, for which the exact Krylov direction is only recovered to leading order in $\tau$, or not at all. 

The comparison between these ansätze allows us to separate several effects: the role of the Hamiltonian structure of the variational manifold, the role of circuit depth and expressivity, and the difference between a well-conditioned subspace and a physically useful Krylov-like subspace. The Krylov and Lanczos references used for benchmarking are not ansätze and are therefore described separately in the numerical protocol.

\begin{table}[t]
\caption{Variational transformations $\mathcal{A}(\bm{\theta})$ considered in this work.
$N_q$ is the number of qubits, $N_P$ the number of Pauli strings entering the qubit representation of $H$, and $N_a$ the number of Hamiltonian groups used in the Trotterized decomposition.
Only $H(\theta)$ contains the exact Krylov step $\ket{\phi_n}\propto H\ket{\phi_{n-1}}$ in its variational manifold, as discussed in the text.}
\label{tab:ansatze}
\begin{ruledtabular}
\begin{tabular}{lccc}
Ansatz & $\mathcal{A}(\bm{\theta})$ & Unitary & Params/iter  \\
\hline
$H(\theta)$    & $\sum_a \theta_a P_a$                                & no  & $N_P$    \\
$VHA(\theta)$  & $\prod_a e^{-i\tau\theta_a H_a}$                     & yes & $N_a$    \\
$HEA(\theta)$  & $U_y(\bm{\theta}_1)\,U_{\mathrm{CNOT}}\,U_y(\bm{\theta}_2)$ & yes & $2N_q$   \\
\end{tabular}
\end{ruledtabular}
\end{table}

\section{Model and Numerical Protocol}
\label{sec:model}

\subsection{Model Hamiltonian}
\label{subsec:model_hamiltonian}

We benchmark the proposed variational Krylov construction on a finite interacting fermionic system of Anderson impurity type, defined in a star geometry \cite{PhysRev.124.41}. The system consists of a single interacting impurity site coupled to $N-1$ non-interacting bath sites, so that $N$ denotes the total number of spatial orbitals. After the Jordan--Wigner transformation, the system is encoded on $N_q=2N$ qubits.
The Anderson impurity model is a standard minimal model for local electronic correlations and impurity physics. Related finite Anderson and Hubbard models have been used as benchmarks for variational quantum algorithms~ \cite{cadeStrategiesSolvingFermiHubbard2020,martinSimulatingStronglyInteracting2022,jamet2021krylovvariationalquantumalgorithm}.

The Hamiltonian is written as
\begin{equation}
    H = H_{\mathrm{1b}} + H_{\mathrm{int}},
\end{equation}
where the one-body term takes the form
\begin{equation}
    H_{\mathrm{1b}} = \sum_{i,\sigma} \epsilon_i\, c^\dagger_{i\sigma} c_{i\sigma}
    + \sum_{i>0,\sigma} V_i \left(c^\dagger_{0\sigma} c_{i\sigma} + \mathrm{h.c.}\right),
\end{equation}
with \(i=0\) denoting the impurity site and \(i>0\) the bath sites. The parameters \(\epsilon_i\) and \(V_i\) respectively correspond to on-site energies and hybridization amplitudes.
The interaction term is local and acts only on the impurity,
\begin{equation}
    H_{\mathrm{int}} = U\, n_{0\uparrow} n_{0\downarrow},
\end{equation}
where \(U\) is the interaction strength.
The fermionic Hamiltonian is mapped to a qubit representation using a Jordan--Wigner transformation, and all calculations are performed in the resulting spin basis.

The star parameters used in this work are obtained from a one-dimensional chain representation. We start from an open chain of $N$ sites with uniform nearest-neighbor hopping $t$, with the impurity at one end and an impurity level $\epsilon_0=-U/2$, and we diagonalize the non-interacting bath block. This yields the star geometry with
\begin{equation}
\epsilon_k = 2t\cos\!\left(\frac{k\pi}{N}\right),
\qquad
V_k = t\sqrt{\frac{2}{N}}\,\sin\!\left(\frac{k\pi}{N}\right),
\label{eq:star_parameters}
\end{equation}
for $k=1,\dots,N-1$. With $\epsilon_0=-U/2$ and a bath spectrum symmetric about zero, the model is particle-hole symmetric at half filling. We set $t=1$, which fixes the energy unit throughout.
This star geometry is particularly convenient for our purposes because it provides a simple and transparent construction of the initial reference state. In the site occupation basis, the bath orbitals are characterized by onsite energies $\epsilon_i$, which can be ordered energetically. We use this ordering to define a Hartree--Fock-like product reference by occupying the lowest-energy spin-orbitals compatible with the chosen particle number and spin sector. After the Jordan--Wigner transformation, this state corresponds to a computational-basis bitstring and can be prepared simply by applying $X$ gates to the occupied spin-orbitals. This provides an easily preparable initial state with the desired quantum numbers, while retaining the impurity-bath hybridization and local interaction in the Hamiltonian used for the subsequent calculations. For $N=4$ at half filling, the four lowest spin-orbitals are the impurity level and the bath level at $\epsilon=-\sqrt{2}\,t$, both doubly occupied.

\subsection{Numerical procedure}
\label{subsec:optimization}

Calculations are initialized from a Hartree--Fock-like product state $\ket{\phi_0}$ with fixed particle number, corresponding to half-filling in the systems considered below, it correspond on the quantum circuit to X gates to the occupied levels.

All calculations reported in this work are performed with exact statevector simulation. Overlaps and Hamiltonian matrix elements are evaluated exactly, so that no shot noise or hardware noise is included. The conditioning analysis presented below therefore isolates the intrinsic near-linear dependence of the generated basis from the additional sampling errors that would be present on a quantum processor, where small eigenvalues of $S$ amplify estimation errors~\cite{epperlyTheoryQuantumSubspace2022,getelinaQuantumSubspaceExpansion2024a}.

Exact reference results are obtained by full diagonalization of the Hamiltonian matrix. The exact ground-state energy and wavefunction are used to evaluate the accuracy of the variational subspace construction. In addition, all variational results are compared to Krylov and Lanczos references. The energy convergence is benchmarked against the exact Lanczos construction, while the conditioning is analyzed using the  normalized Krylov basis without reorthogonalization. This distinction makes explicit the role played by orthogonalization in the standard Lanczos algorithm and the conditioning issues that arise in  Krylov subspaces.

For each ansatz, the parameters are optimized at each iteration by minimizing the cost function defined in Eq.~\eqref{eq:variational_lanczos_cost}. The optimization is performed using the COBYLA nonlinear optimizer, with a maximum of $10000$ optimization steps per variational update.
Variational quantum circuit optimization is known to depend strongly on initialization, ansatz structure, parameter redundancy, and landscape properties~\cite{mccleanBarrenPlateausQuantum2018,grantInitializationStrategyAddressing2019,bittelTrainingVariationalQuantum2021,laroccaBarrenPlateausVariational2025,choyMolecularEnergyLandscapes2023}.

It should also be noted that the cost of a single variational update is not dominated by the classical linear algebra, but by the number of matrix elements to be evaluated. At iteration $n$, the blocks of $H^{VQKD}$ and $S$ associated with the previously optimized states do not depend on $\bm{\theta}_n$ and are evaluated only once, so that each evaluation of $\mathcal{C}_n$ requires the $n$ overlaps $\braket{\phi_i}{\phi_n}$ and the $n+1$ elements $\bra{\phi_i}H\ket{\phi_n}$ associated with the trial state, the regularization term reusing the same overlaps at no additional cost. A full iteration therefore involves $\mathcal{O}(n\,N_{\mathrm{opt}})$ such estimations, with $N_{\mathrm{opt}}$ the number of optimizer evaluations, and the construction of a subspace of dimension $D$ requires $\mathcal{O}(D^{2}N_{\mathrm{opt}})$. The factor $N_{\mathrm{opt}}$ is the intrinsic overhead of the variational search, since a non-variational quantum Krylov construction requires only the $\mathcal{O}(D^{2})$ distinct matrix elements. This estimate counts matrix-element evaluations, not the shots or state-preparation gates required for them. On a quantum processor, the resulting measurement and optimization overhead could be a major limitation of the construction in its present form.

Unless otherwise specified, the optimization is initialized from the Hamiltonian-term coefficients when such an initialization is available. This choice is used for the Hamiltonian-inspired ansatz, for which the variational parameters are naturally associated with the operator decomposition of the Hamiltonian. For the HEA ansatz, where no such Hamiltonian-based initialization exists, random initializations and a small number of restarts can be used to improve robustness. The number of variational parameters depends on the ansatz considered, as listed in Table~\ref{tab:ansatze}. For the Hamiltonian-inspired ansatz it is set by the number of Pauli terms in the qubit representation of $H$.

The full numerical procedure is summarized in Algorithm~\ref{alg:vkc}. At each iteration, a trial state is generated from the current ansatz and optimized according to Eq.~\eqref{eq:variational_lanczos_cost}. During the optimization, the projected Hamiltonian, overlap matrix, and associated orthonormal representation are updated for each trial set of parameters. Once the minimum is reached, the optimized state is added to the basis and the generalized eigenvalue problem is solved in the enlarged variational subspace.

\begin{algorithm}[ht]
\caption{Variational Krylov construction}
\label{alg:vkc}

\begin{enumerate}

\item \textbf{Initialization}:
\begin{enumerate}
\item Choose an initial normalized reference state
\begin{equation*}
\ket{\phi_0} = \ket{\Phi_0}.
\end{equation*}

\item Initialize the variational basis as
\begin{equation*}
\mathcal{V}_0 = \mathrm{span}\{\ket{\phi_0}\}.
\end{equation*}
\end{enumerate}

\item \textbf{Iterative construction}:

For $n = 1,\dots,N_{\rm iter}$:

\begin{enumerate}

\item Optimize $\bm{\theta}_n$ by minimizing $\mathcal{C}_n(\bm{\theta}_n)$, Eq.~\eqref{eq:variational_lanczos_cost}. Each evaluation of the cost function proceeds as follows:

\begin{enumerate}
\item Generate and normalize the trial state
\begin{equation*}
\ket{\phi_n(\bm{\theta}_n)}
\propto
\mathcal{A}(\bm{\theta}_n)\ket{\phi_{n-1}}.
\end{equation*}

\item Keep the previously optimized states $\{\ket{\phi_i(\bm{\theta}_i^*)}, i<n\}$ fixed during the cost evaluation.

\item Update the projected matrices
\begin{equation*}
H^{VQKD}_{ij}
=
\bra{\phi_i} H \ket{\phi_j},
\qquad
S_{ij}
=
\braket{\phi_i}{\phi_j},
\end{equation*}
for $0\leq i,j\leq n$.

\item Construct the orthonormal representation
\begin{equation*}
S = LL^\dagger ,
\qquad
H^{\perp}
=
L^{-1}H^{VQKD} L^{-\dagger} ,
\end{equation*}
and read off $H^{\perp}_{n,n-1}$.
\end{enumerate}

\item Add the optimized state $\ket{\phi_n(\bm{\theta}_n^*)}$ to the variational basis.

\item Solve the generalized eigenvalue problem in the enlarged subspace,
\begin{equation*}
H\bm{c}
=
E S \bm{c}.
\end{equation*}

\end{enumerate}

\end{enumerate}

\end{algorithm}

\section{Results and Discussion}
\label{sec:results_discussion}

All results in this section are obtained for the $N=4$ Anderson impurity system at half filling which corresponds to 8 qubits calculations. We assess the variational construction using two main observables. The first one is the relative ground-state energy error,
\begin{equation}
\epsilon_E^{(n)}
=
1-\frac{E^{(n)}}{E_0}.
\label{eq:relative_energy_error}
\end{equation}
Here, $E^{(n)}$ is the lowest eigenvalue obtained from the generalized eigenvalue problem in the variational subspace after $n$ iterations, and $E_0$ is the exact ground-state energy. Since the exact ground-state energy is negative for the systems considered here, this convention gives a positive relative error that vanishes when $E^{(n)}=E_0$.

The second observable is the condition number of the overlap matrix, $\kappa(S)$, defined in Eq.~\eqref{eq:condition_number}. This quantity measures the numerical stability of the non-orthogonal basis. A large value of $\kappa(S)$ indicates the presence of nearly linearly dependent states and therefore a potentially unstable generalized eigenvalue problem.

In the figures below, curves that leave the lower edge of a panel do so because of the vertical range shown, which does not necessarily mean that the corresponding quantity has converged; the values reached at the largest iteration are collected in Table~\ref{tab:ansatz_performance}.

\subsection{Role of the regularization parameter}
\label{subsec:lambda_results}

\begin{figure}[t]
\centering
\includegraphics[width=\columnwidth]{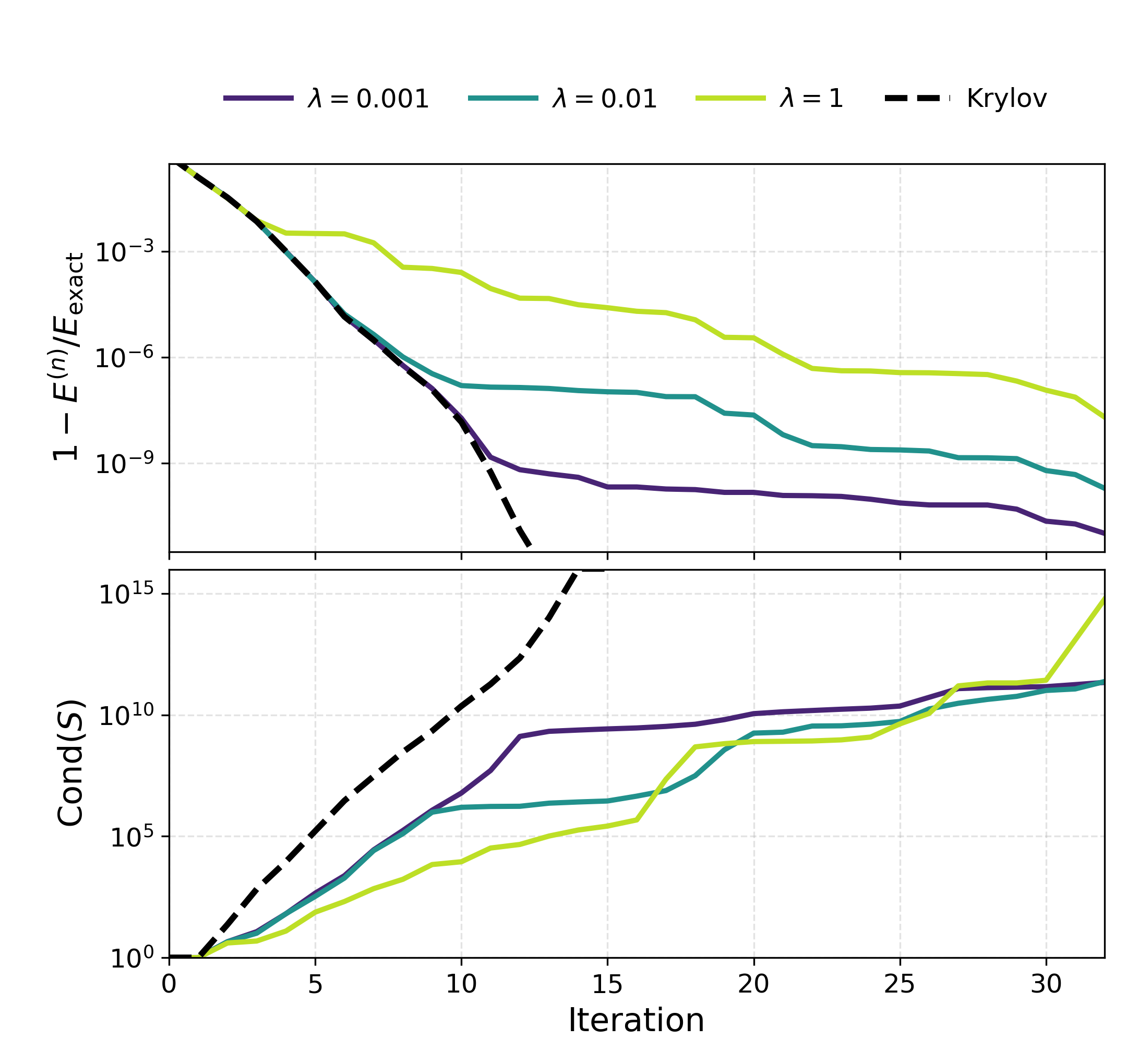}
\caption{
Role of the regularization parameter $\lambda$ for the $H(\theta)$ ansatz, shown for $\lambda=10^{-3}$, $10^{-2}$ and $1$.
Top panel: relative ground-state energy error $\epsilon_E^{(n)}$ as a function of the iteration number.
Bottom panel: condition number $\kappa(S)$ of the overlap matrix.
The black dashed curve is the exact Krylov reference, which coincides with the $\lambda=0$ limit of this ansatz (see Sec.~\ref{subsec:ansatz_constructions}); in the upper panel it leaves the frame once the energy error reaches numerical precision.
Increasing $\lambda$ degrades the energy convergence monotonically, whereas the condition number remains within one to two orders of magnitude across the three values, while staying far below the  Krylov curve.
Data shown for the $N=4$, $U=4$ system.
}
\label{fig:lambda_tradeoff}
\end{figure}

We first analyze the role of the regularization parameter $\lambda$, which controls the competition between following Hamiltonian-connected Krylov-like directions and suppressing near-linear dependencies in the generated basis. Unless stated otherwise, this analysis is performed using the $H(\theta)$ ansatz.
Figure~\ref{fig:lambda_tradeoff} shows the relative energy error and the condition number of the overlap matrix for $\lambda=10^{-3}$, $10^{-2}$ and $1$. As discussed in Sec.~\ref{subsec:ansatz_constructions}, the exact Krylov reference corresponds to the $\lambda=0$ limit of this ansatz, so that the deviations from the dashed curve measure the effect of the overlap penalty alone.

Two observations follow. First, the effect of the regularization on the energy convergence is monotonic and pronounced. For $\lambda=10^{-3}$ and $10^{-2}$, the construction follows the exact Krylov convergence over the first ten iterations and then enters a much slower regime, reaching $\epsilon_E\simeq 10^{-9}$ and $10^{-7}$ around $n\simeq 15$ and improving only marginally over the remaining iterations. For $\lambda=1$, the construction departs from the Krylov curve after only a few iterations and the energy error remains several orders of magnitude higher over the whole range of iterations considered.

Second, and in contrast, the conditioning is only weakly sensitive to $\lambda$ in this range. All three regularized constructions keep $\kappa(S)$ several orders of magnitude below the  Krylov value, which grows exponentially and exceeds $10^{15}$ within about fifteen iterations. However, the three curves remain within one to two orders of magnitude of each other.

Taken together, these results indicate that, at least in the range explored here, $\lambda$ does not act as a smooth knob trading convergence against conditioning. The bulk of the conditioning gain is already obtained at very small $\lambda$: it is the presence of the penalty, rather than its magnitude, that prevents the collapse observed for the  Krylov basis. Increasing $\lambda$ beyond this point costs convergence without buying a corresponding stabilization, because the optimization is then driven toward directions that are nearly orthogonal to the existing basis but not the most relevant ones for the low-energy subspace. A weak penalty is therefore preferable, and the interesting question is not how large $\lambda$ should be, but whether a small, possibly iteration-dependent penalty is sufficient to control the conditioning over longer recursions. Adaptive or iteration-dependent regularization schedules are a natural extension in this direction, but they are not explored systematically in the present work.

\subsection{Role of the ansatz structure}
\label{subsec:ansatz_results}

\begin{figure}[t]
\centering
\includegraphics[width=\columnwidth]{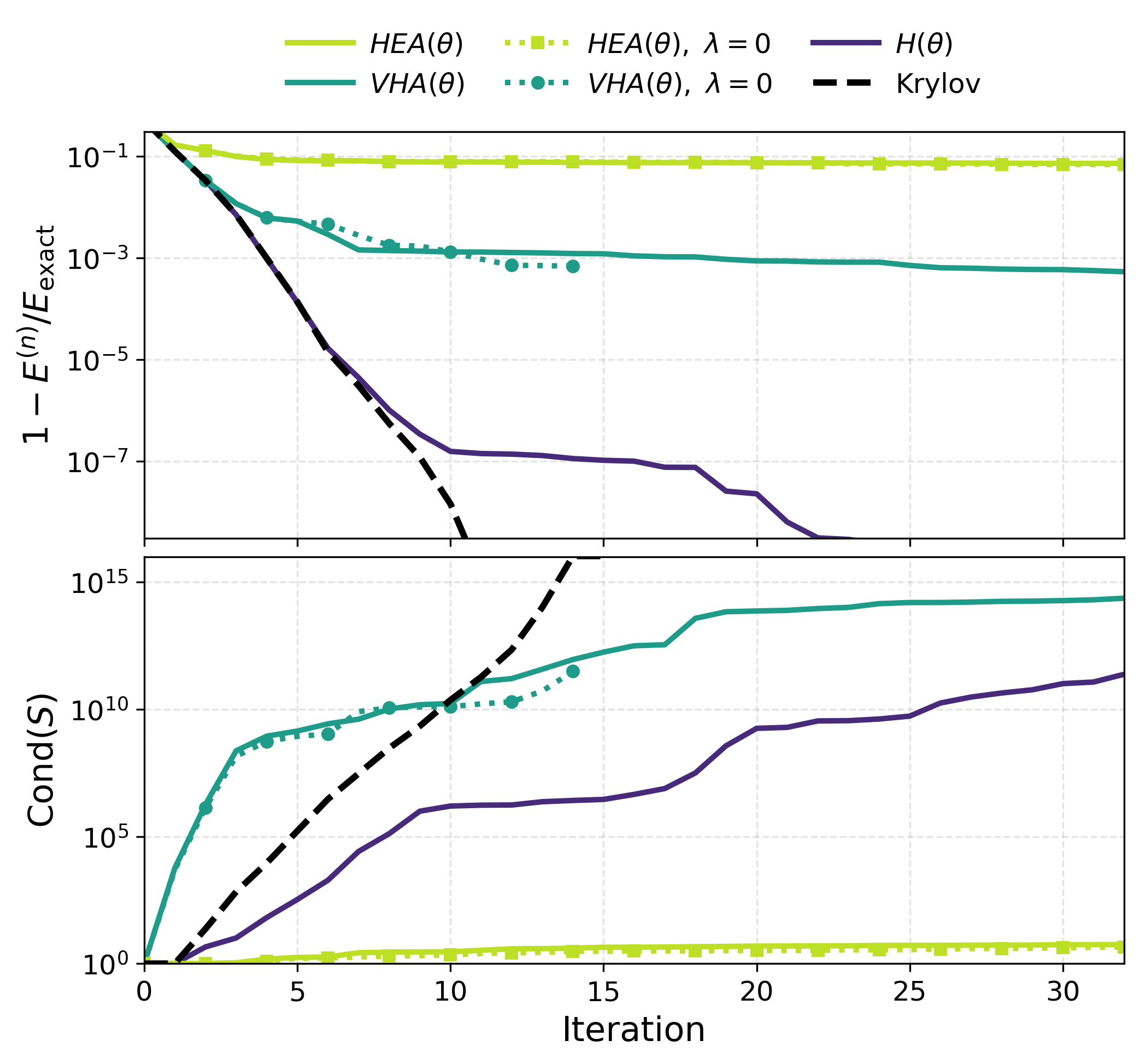}
\caption{
Comparison of ansatz structures for the $N=4$, $U=4$ Anderson impurity system.
Unless explicitly indicated, the variational curves are obtained at fixed regularization $\lambda=0.01$.
The dotted curves show additional $VHA(\theta)$ and $HEA(\theta)$ runs with $\lambda=0$; the $VHA(\theta)$ run at $\lambda=0$ terminates at $n\simeq14$, where the Cholesky factorization of the overlap matrix fails.
The upper panel shows the relative ground-state energy error $\epsilon_E^{(n)}$, while the lower panel shows the condition number $\kappa(S)$ of the overlap matrix. The $H(\theta)$ curve leaves the upper panel around $n\simeq22$ because of the vertical range shown; its value at larger $n$ is given in Table~\ref{tab:ansatz_performance}.
The curve labelled ``Krylov'' is the reference obtained from the  Krylov basis: in the upper panel it is the energy convergence of the corresponding (exact Lanczos) subspace, and in the lower panel the conditioning of the  non-orthogonal basis itself.
The comparison shows that the dominant limitation of the shallow $HEA(\theta)$ and $VHA(\theta)$ curves is mainly due to ansatz structure rather than to the overlap regularization alone.
}
\label{fig:ansatz_U4}
\end{figure}

We now compare different variational ansätze in the representative interacting case $U=4$. The main comparison is performed at fixed regularization $\lambda=0.01$, which allows us to analyze the effect of the ansatz structure at a fixed cost-function balance. In addition, for the $VHA(\theta)$ and $HEA(\theta)$ ansatz constructions, we include extra calculations at $\lambda=0$, shown as dotted curves in Fig.~\ref{fig:ansatz_U4}. These additional curves are used as a control to assess whether the observed behavior is mainly caused by the overlap regularization or by the variational manifold itself. We compare the Hamiltonian-inspired update $H(\theta)$, the Trotterized variational Hamiltonian ansatz $VHA(\theta)$, and the single-entangling-layer hardware-efficient ansatz $HEA(\theta)$.
Figure~\ref{fig:ansatz_U4} shows the relative ground-state energy error and the condition number of the overlap matrix. The exact Krylov construction provides the fastest energy convergence, but it rapidly produces an extremely ill-conditioned overlap matrix. In contrast, the variational constructions deviate from exact Krylov behavior at larger iteration numbers while maintaining a significantly better conditioned basis.
A clear hierarchy emerges between the different ansatz constructions. The Hamiltonian-inspired update $H(\theta)$ systematically provides the fastest variational convergence. It follows the Krylov curve closely during the first iterations and then enters a slower, regularized convergence regime. This behavior indicates that the Hamiltonian-informed manifold is well aligned with the dominant Krylov-like directions at early depths, while avoiding the severe conditioning collapse of the Krylov basis.

The behavior of the $VHA(\theta)$ construction is intermediate. It is physically motivated and improves the energy over many iterations, but for the choice $\tau=0.1$ used here it does not reproduce the rapid initial Krylov-like convergence as efficiently as the Hamiltonian-recursive update. The additional $\lambda=0$ curve shows that removing the overlap regularization does not qualitatively change this conclusion. It even makes the procedure unstable: without the penalty, the generated states become nearly linearly dependent within a few iterations, the smallest eigenvalue of $S$ drops below the numerical resolution, and the Cholesky factorization of the overlap matrix fails, which terminates the recursion at $n\simeq 14$. The $VHA(\theta)$ performance is therefore not primarily limited by the regularization term, but by the structure and time scale of the ansatz itself. A smaller value of $\tau$ keeps the generated states closer to the previous state, while a larger value of $\tau$ allows larger moves in Hilbert space but weakens the direct connection with the leading-order Krylov-like expansion.

The $HEA(\theta)$ remains extremely well conditioned but exhibits poor energy convergence. This result should not be interpreted as a general statement about all hardware-efficient circuits, since it depends on the number of layers and on the entangling geometry. In the present work, the HEA is deliberately restricted to two layers of real $R_y$ rotations separated by a single CNOT cascade, and therefore provides a shallow baseline. The $\lambda=0$ curve confirms that its poor convergence is not caused by an excessive overlap penalty. Even without regularization, the HEA remains well conditioned and does not efficiently capture the low-energy Hamiltonian-generated directions. Good conditioning alone is therefore not a sufficient criterion for a useful Krylov-like subspace. More layers would be required at the cost of a more difficult optimization to improve the results.

Table~\ref{tab:ansatz_performance} summarizes this comparison quantitatively at two representative depths.
It makes the trade-off explicit: the two quantities that a useful construction must control simultaneously, the energy error and the conditioning, are optimized by different ansätze, and only $H(\theta)$ achieves a reasonable compromise between them.

\begin{table}[t]
\caption{Performance of the ansatz constructions for the $N=4$, $U=4$ system at
$\lambda=0.01$, compared with the exact Krylov reference. $\epsilon_E^{(n)}$ is
the relative ground-state energy error of Eq.~\eqref{eq:relative_energy_error}
and $\kappa(S)$ the condition number of the overlap matrix, both evaluated at
iteration $n=10$ and $n=32$. Entries marked with a dash indicate that the
quantity has reached numerical precision, or that $\kappa(S)$ has grown beyond
the range where the projected problem remains meaningful.}
\label{tab:ansatz_performance}
\begin{ruledtabular}
\begin{tabular}{lcccc}
 & \multicolumn{2}{c}{$\epsilon_E^{(n)}$} & \multicolumn{2}{c}{$\kappa(S)$} \\
\cline{2-3}\cline{4-5}
Construction & $n=10$ & $n=32$ & $n=10$ & $n=32$ \\
\hline
Krylov (exact)  & $1.4\times10^{-8}$ & ---                  & $10^{10}$ & --- \\
$H(\theta)$     & $1.5\times10^{-7}$ & $1.9\times10^{-10}$  & $10^{6}$  & $10^{11}$ \\
$VHA(\theta)$   & $1.2\times10^{-3}$ & $5.3\times10^{-4}$   & $10^{10}$ & $10^{14}$ \\
$HEA(\theta)$   & $7.6\times10^{-2}$ & $7.1\times10^{-2}$   & $2.9$     & $5.5$ \\
\end{tabular}
\end{ruledtabular}
\end{table}

Overall, Fig.~\ref{fig:ansatz_U4} shows that the ansatz structure is a central ingredient of the construction. The additional $\lambda=0$ curves indicate that, for the shallow $VHA(\theta)$ and $HEA(\theta)$ considered here, removing the overlap regularization is not enough to recover Krylov-like convergence. The most successful variational directions are not selected by generic expressivity alone, but by the compatibility between the ansatz manifold and the operator content of the Hamiltonian. The variational construction should therefore be understood as a structured compromise between Krylov-like convergence, conditioning control, and optimization feasibility.

\subsection{Stability with respect to the initialization}
\label{subsec:optimization_stability}

We next investigate the stability of the variational optimization with respect to the choice of initial parameters. The previous results show that the method initially follows a Krylov-like convergence before entering a slower, regularized regime. A natural question is whether this saturation is only due to the limited expressivity of the ansätze, or whether the optimization itself becomes increasingly sensitive as the iteration number grows.

\begin{figure}[t]
\centering
\includegraphics[width=\columnwidth]{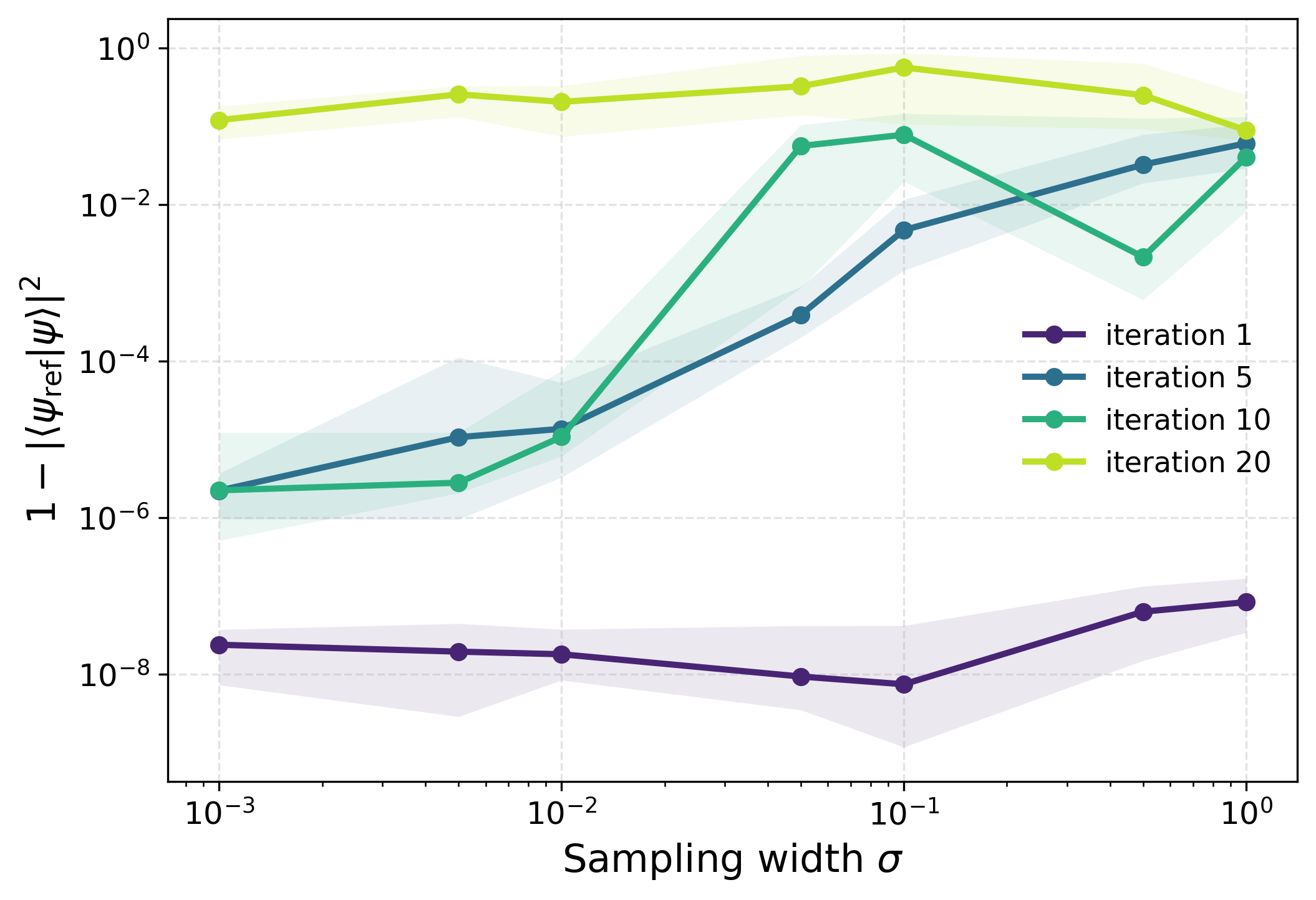}
\caption{
Stability of the variational optimization with respect to the initialization.
For each iteration, the initial parameter vector is perturbed by Gaussian noise of width $\sigma$, and the optimization is repeated.
The plotted quantity is the median infidelity between the resulting optimized state and the reference state obtained from the unperturbed initialization, while the shaded regions indicate the interquartile range.
Early iterations display a robust attraction toward the same optimized state, whereas later iterations show a strong increase of the dispersion.
This indicates increasing sensitivity to initialization under the fixed optimizer and finite optimization budget used in this work.
Data shown for the $N=4$, $U=4$ system using the $H(\theta)$ ansatz at $\lambda=0.01$.
}
\label{fig:basin_stability}
\end{figure}

To probe this effect, we perform a perturbative analysis of the initial parameters. For a given iteration $n$, we first obtain a reference optimized state $\ket{\psi_{\mathrm{ref}}}$ from the unperturbed initialization. We then perturb the initial parameter vector by adding Gaussian noise of width $\sigma$, and rerun the optimization starting from this perturbed initialization. The final optimized state $\ket{\psi}$ is compared to the reference optimized state through the infidelity
\begin{equation}
\delta_{\mathrm{inf}}
=
1
-
\left|
\braket{\psi_{\mathrm{ref}}}{\psi}
\right|^2 .
\label{eq:infidelity_basin}
\end{equation}
The median value over different perturbed initializations is shown in Fig.~\ref{fig:basin_stability}, while the shaded regions indicate the interquartile range.

At the first iteration, the optimized state is extremely robust with respect to the initialization. The dispersion remains close to numerical precision over the full range of perturbation widths considered, indicating that the optimization flows toward a well-defined direction. At intermediate iterations, the situation changes qualitatively. The final state remains stable for very small perturbations of the initial parameters, but the dispersion increases once $\sigma$ becomes large enough. At later iterations, the dispersion becomes large even for moderate perturbations, indicating a stronger dependence on the initial point in parameter space.

This analysis should be interpreted with some caution. The observed sensitivity is not a direct proof of a fragmented optimization landscape, because it may also depend on the optimizer, the finite COBYLA iteration budget, and the parameterization of the ansatz. In particular, a larger number of optimization steps or a different nonlinear optimizer could modify the quantitative behavior. Nevertheless, under the fixed optimization protocol used throughout this work, the trend is clear: early iterations are robust, whereas later iterations become increasingly non-unique at the level of the optimized wave function.

This non-uniqueness does not necessarily imply a failure of the variational subspace construction. Once the dominant low-energy directions have already been captured, several distinct variational states may produce comparable changes in the projected energy or in the conditioning of the basis.

This observation also clarifies why state-by-state overlaps with Lanczos vectors are not sufficient to characterize the method. A variational state may differ strongly from the corresponding Lanczos vector while still contributing to a subspace that captures similar low-energy information. This motivates the subspace-level comparison discussed next.

\subsection{Relation to the Lanczos subspace}
\label{subsec:subspace_lanczos}

The preceding analysis shows that the optimized states can become non-unique at the level of wave-function overlap. This suggests that a state-by-state comparison with Lanczos vectors is not the most appropriate diagnostic. Different bases may span nearly the same physically relevant subspace while having small overlaps between individual basis vectors.
To assess the relation between the variational construction and Lanczos at the subspace level, we compare the cumulative VQKD subspace
\begin{equation}
\mathcal{V}_{\mathrm{VQKD}}(D)
=
\operatorname{span}
\left\{
\ket{\phi_0},
\ket{\phi_1},
\ldots,
\ket{\phi_{D-1}}
\right\}
\label{eq:vqkd_subspace}
\end{equation}
with a fixed Lanczos reference subspace
\begin{equation}
\mathcal{K}_{D_L}
=
\operatorname{span}
\left\{
\ket{q_0},
\ket{q_1},
\ldots,
\ket{q_{D_L-1}}
\right\}.
\label{eq:lanczos_reference_subspace}
\end{equation}
In practice, both sets are first converted into orthonormal bases. We then compute the singular values of the overlap matrix between these two orthonormal bases. These singular values are the cosines of the principal angles $\theta_k$ between the two subspaces.

For each VQKD dimension $D$, we define the dimension-normalized subspace mismatch
\begin{equation}
\Delta_D
=
1
-
\frac{1}{r}
\sum_{k=1}^{r}
\cos^2 \theta_k ,
\qquad
r = \min(D,D_L).
\label{eq:principal_angle_mismatch}
\end{equation}
This normalization removes the trivial dimension mismatch between $\mathcal{V}_{\mathrm{VQKD}}(D)$ and $\mathcal{K}_{D_L}$, but its interpretation depends on which subspace is smaller. For $D\leq D_L$, $\Delta_D=0$ means that the VQKD subspace is contained in the Lanczos reference subspace, whereas for $D\geq D_L$ it means that the Lanczos reference subspace is contained in the VQKD subspace. The quantity therefore measures subspace capture rather than a one-to-one correspondence between basis vectors.

\begin{figure}[t]
\centering
\includegraphics[width=\columnwidth]{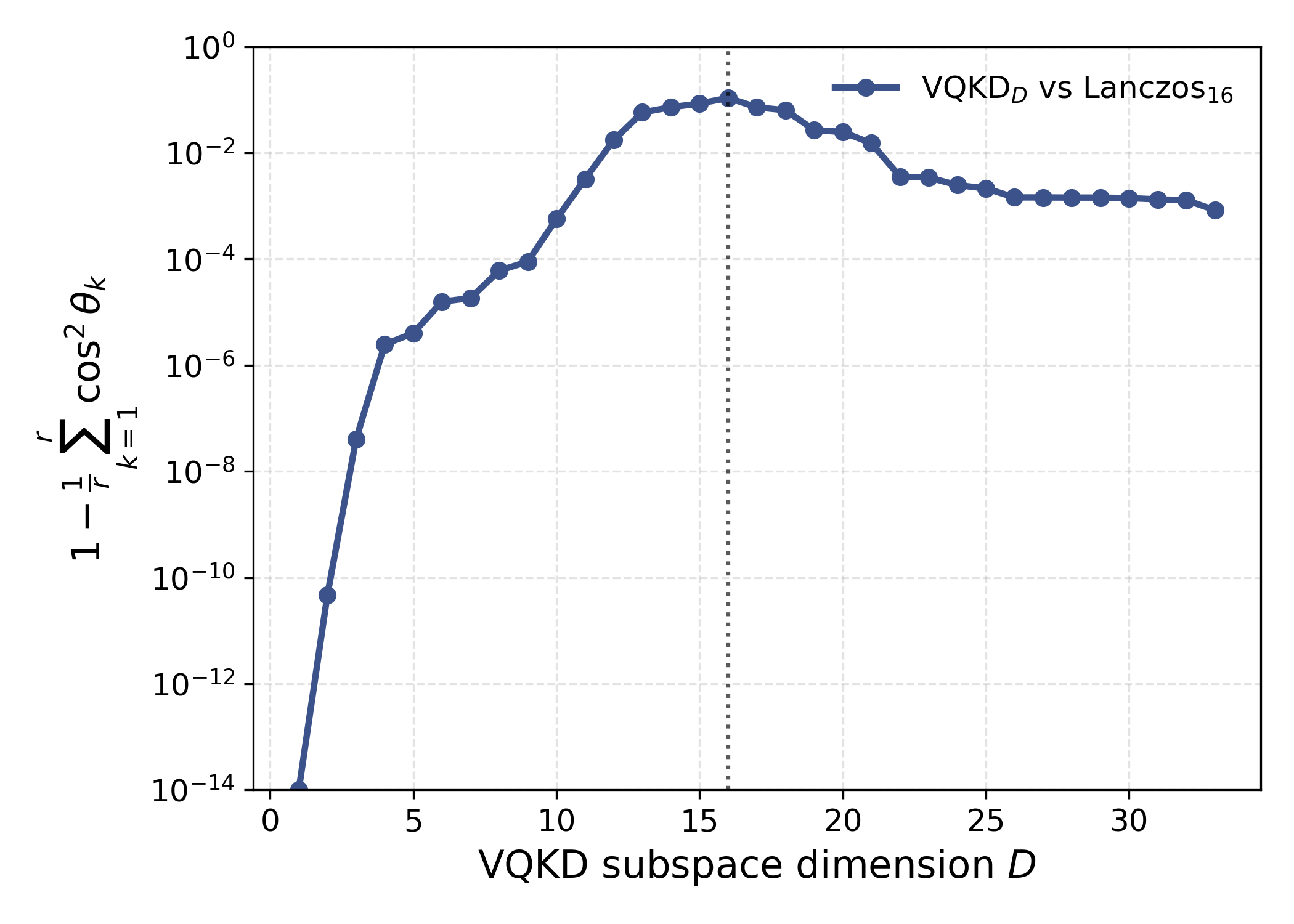}
\caption{
Principal-angle mismatch between the cumulative VQKD subspace $\mathcal{V}_{\mathrm{VQKD}}(D)$ and a fixed Lanczos reference subspace $\mathcal{K}_{D_L}$ with $D_L=16$.
The plotted quantity is $\Delta_D = 1-r^{-1}\sum_{k=1}^{r}\cos^2\theta_k$, where $\theta_k$ are the principal angles between the two subspaces and $r=\min(D,D_L)$.
Small values indicate that the VQKD subspace captures the Lanczos reference subspace as well as allowed by its dimension.
The vertical dotted line marks the reference dimension $D_L=16$.
The increase at intermediate dimensions reflects the deviation from the fixed Lanczos reference space. For $D>D_L$, the decrease is expected from the nested cumulative VQKD subspaces; the saturation around $\Delta_D\simeq10^{-3}$ quantifies the residual average mismatch with the fixed Lanczos subspace.
Data shown for the $N=4$, $U=4$ system using the $H(\theta)$ ansatz at $\lambda=0.01$.
}
\label{fig:lanczos_subspace_capture}
\end{figure}

Figure~\ref{fig:lanczos_subspace_capture} shows this quantity for $D_L=16$. At very small $D$, the mismatch is close to numerical precision, showing that the early VQKD subspace lies almost entirely within the fixed Lanczos reference space. As $D$ increases, the mismatch grows, indicating that the variational construction progressively explores directions outside this reference subspace. Around $D=D_L$, the mismatch is substantial, confirming that the two equal-dimensional subspaces are not identical.

For $D>D_L$, $\Delta_D$ decreases because the cumulative VQKD subspaces are nested and can only capture more of the fixed Lanczos reference space as their dimension grows. The relevant observation is therefore the value reached: the mismatch saturates around $\Delta_D\simeq10^{-3}$, corresponding to a mean squared principal-angle overlap of about $0.999$. Thus, although the variational construction does not reproduce the Lanczos basis at equal dimension, an enlarged VQKD subspace captures the fixed Lanczos reference subspace with high average accuracy.

This result reinforces the main interpretation of the numerical study. The method should not be viewed as an exact variational implementation of the Lanczos recursion. Rather, it constructs a regularized Krylov-like subspace whose relation to Lanczos is approximate and dimension dependent. The relevant object is therefore the generated subspace, not the individual optimized states.

\section{Conclusion}
\label{sec:conclusion}

We have investigated a variational Krylov-like construction that replaces the direct Lanczos recurrence by optimized state transformations while penalizing near-linear dependencies. On the small Anderson impurity benchmark, a weak overlap penalty is already sufficient to delay the severe conditioning collapse of the Krylov basis, whereas stronger regularization mainly slows the energy convergence. The dominant factor is instead the structure of the variational manifold: the Hamiltonian-inspired $H(\theta)$ update best preserves the early Krylov-like convergence, while the shallow $HEA(\theta)$ remains well conditioned but fails to capture useful low-energy directions. The principal-angle analysis further shows that the optimized states need not reproduce Lanczos vectors individually: although the equal-dimensional subspaces eventually differ, an enlarged VQKD subspace recaptures the fixed Lanczos reference space with a mean mismatch of order $10^{-3}$. The construction is therefore better viewed as a regularized Krylov-like subspace than as a variational realization of Lanczos itself.

From a quantum-algorithm perspective, this construction avoids implementing explicit coherent Lanczos orthogonalization, at the expense of variational optimization and overlap measurements. The present study remains limited to a small exact-statevector benchmark and does not include sampling noise, while the $\mathcal{O}(D^{2}N_{\mathrm{opt}})$ measurement cost would be substantial on hardware. A natural perspective is therefore to combine the variational and recursive viewpoints. Short recursive Lanczos-like blocks could retain the local Krylov structure, while periodic variational compression or re-preparation could prevent the recursive state-preparation circuit from growing indefinitely. Conversely, the recursively generated directions could provide structured targets that reduce ansatz bias and optimization ambiguity. Whether such a hybrid quantum-classical construction can simultaneously control conditioning, optimization cost, and recursive circuit growth is an open question and a promising direction for future work.

\begin{acknowledgments}
This work was supported in part by the Maison du Quantique de Nouvelle-Aquitaine “HybQuant”, as part of the
HQI initiative and France 2030, under the French National Research Agency (ANR) grant ANR-22-PNCQ-0002 and the PEPR Quantique under Grant No. ANR-23-PETQ-0006.
\end{acknowledgments}

\appendix

\section{Orthogonalization procedure and recursive construction of the effective Hamiltonian}
\label{app:orthogonalization}

This appendix collects the recursive formulas used to update the orthonormal representation $H^{\perp}$ of Eq.~\eqref{eq:hperp} when a single state is added to the basis, without ever forming the orthonormalized states explicitly. This is the update actually performed at every evaluation of the cost function in Algorithm~\ref{alg:vkc}.

\subsection{Partitioned form and orthonormalization of the previous subspace}

Given a sequence of generally non-orthogonal states $\{\ket{\phi_0},\dots,\ket{\phi_k}\}$ with projected matrices $H^{VQKD}$ and $S$ defined as in Eq.~\eqref{eq:projected_matrices}, it is convenient at iteration \(k\) to partition
\begin{equation}
    H^{VQKD} =
    \begin{pmatrix}
        H_o & h \\
        h^\dagger & H_x
    \end{pmatrix},
    \qquad
    S =
    \begin{pmatrix}
        S_o & b \\
        b^\dagger & S_x
    \end{pmatrix},
\end{equation}
where the subscript \(o\) denotes the previously constructed subspace, and the vectors \(h\) and \(b\) encode the couplings to the newly added state \(\ket{\phi_k}\).

The previous subspace is orthonormalized through a Cholesky factorization of its overlap block,
\begin{equation}
    S_o = L L^\dagger,
\end{equation}
where \(L\) is a lower triangular matrix. Defining transformed states
\begin{equation}
    \ket{\chi_i} = \sum_j \ket{\phi_j} (L^{-\dagger})_{ji},
\end{equation}
one obtains an orthonormal basis satisfying \(\braket{\chi_i }{ \chi_j} = \delta_{ij}\), in which the projected Hamiltonian reads
\begin{equation}
    H^{\perp}_{\mathrm{prev}} = L^{-1} H_o L^{-\dagger}.
\end{equation}

\subsection{Recursive update for the new state}

The overlap and Hamiltonian couplings of the new state are transformed as
\begin{equation}
    b_t = L^{-1} b,
    \qquad
    h_t = L^{-1} h.
\end{equation}

The squared norm of the component orthogonal to the previous subspace is given by
\begin{equation}
    s_\perp = S_x - b_t^\dagger b_t.
\end{equation}
Since all variational states are normalized, \(S_x = 1\) and
\begin{equation}
    s_\perp = 1 - b_t^\dagger b_t .
\end{equation}
This quantity vanishes when the new state becomes linearly dependent on the existing basis, and its numerical breakdown is what terminates the unregularized recursion discussed in Sec.~\ref{subsec:ansatz_results}.

The orthonormalized Hamiltonian is then updated recursively. The off-diagonal coupling between the new state and the previous subspace is given by
\begin{equation}
    h_{\mathrm{new}} =
    \frac{
        h_t - H^{\perp}_{\mathrm{prev}} b_t
    }{
        \sqrt{s_\perp}
    },
\end{equation}
and the corresponding diagonal matrix element reads
\begin{equation}
    H^{\perp}_{kk}
    =
    \frac{
        H_x
        - h_t^\dagger b_t
        - b_t^\dagger h_t
        + b_t^\dagger H^{\perp}_{\mathrm{prev}} b_t
    }{
        s_\perp
    }.
\end{equation}

The full orthonormal Hamiltonian is therefore
\begin{equation}
    H^{\perp} =
    \begin{pmatrix}
        H^{\perp}_{\mathrm{prev}} & h_{\mathrm{new}} \\
        h_{\mathrm{new}}^\dagger & H^{\perp}_{kk}
    \end{pmatrix},
\end{equation}
and $H^{\perp}_{k,k-1}$ is the effective Lanczos-like coupling entering the cost function, Eq.~\eqref{eq:variational_lanczos_cost}.

\section{Connection to Gram-Schmidt orthogonalization}
\label{app:GS}

\subsection{Construction in Appendix \ref{app:orthogonalization}}

The construction  in Appendix \ref{app:orthogonalization} is equivalent to performing an explicit Gram-Schmidt orthogonalization of the states. Indeed, the orthogonal component of \(\ket{\phi_k}\) can be written as
\begin{equation}
    \ket{\phi_k^\perp}
    \propto
    \ket{\phi_k}
    -
    \sum_{i,j<k}
    (S_o^{-1})_{ij} \, b_j \, \ket{\phi_i}.
\end{equation}
The procedure described above reconstructs exactly the Hamiltonian that would be obtained in the orthonormal basis \(\{\ket{\phi_i^\perp}\}\), without ever explicitly forming these states: all required quantities follow from overlaps and Hamiltonian matrix elements between the variational states, which is what makes the scheme compatible with hybrid quantum-classical implementations, as discussed in Sec.~\ref{subsec:iterative_construction}.

\subsection{Quantum Lanczos construction in section \ref{sec:direct_lanczos}}

For completeness, we mention that the construction in section \ref{sec:direct_lanczos} can be rewritten
\begin{equation}
\label{eq:lanczos2}
\begin{split}
| {q}_0 \rangle 
\\
| {q}_1 \rangle 
&\propto 
\Pi_{0,-1} {H} | {q}_0 \rangle
\\
\forall\alpha\ge 1:\quad| {q}_{\alpha+1} \rangle 
&\propto 
\Pi_{\alpha,\alpha-1} H | {q}_\alpha \rangle,
\end{split}
\end{equation}
where
\begin{equation}
\label{eq:GS}
\begin{split}
\Pi_{0,-1} 
&= I - | {q}_0 \rangle \langle {q}_0 | \\
\forall\alpha\ge1:\quad \Pi_{\alpha,\alpha-1} 
&= I - \sum_{\beta=\alpha-1}^{\alpha} | {q}_\beta \rangle \langle {q}_\beta |
, 
\end{split}
\end{equation}
represents the quantum Gram-Schmidt (orthogonalization) operators, which are projectors ($\Pi_{\alpha,\alpha-1}^2 =\Pi_{\alpha,\alpha-1}^\dagger =\Pi_{\alpha,\alpha-1}$ with eigenvalues $0$ or $1$).


\subsubsection{LCU formulation}

We have the linear combination of unitaries (LCU) decomposition
%
%
%
%
\begin{equation}
\begin{split}
\Pi_{\alpha,\alpha-1} = 
\frac{1}{2}I+ \frac{1}{2}R_{\alpha,\alpha-1}
,
\end{split}
\end{equation}
where
\begin{equation}
\begin{split}
R_{\alpha,\alpha-1} =  I-2|{q}_{\alpha-1} \rangle \langle {q}_{\alpha-1} | - 2|{q}_\alpha \rangle \langle {q}_\alpha |
,
\end{split}
\end{equation}
is a (unitary and hermitian) reflection operator ($R_{\alpha,\alpha-1}^2 = I$ and $R_{\alpha,\alpha-1}^\dagger=R_{\alpha,\alpha-1}$).
Considering $H=\sum_{a=0}^{T-1} h_a P_a$ normalized such that $||H||\le 1$, this leads to the LCU decomposition
\begin{equation}
\begin{split}
\Pi_{\alpha,\alpha-1}H = 
\sum_{a=0}^{T-1}\frac{h_a}{2}P_a 
+
\sum_{a=T}^{2T-1}\frac{h_{a-T}}{2}R_{\alpha,\alpha-1}P_{a-T} 
,
\end{split}
\end{equation}
which  satisfies $||\Pi_{\alpha,\alpha-1}H||\le 1$.
We have (because the $\{| {q}_{\alpha}\rangle\}$ are orthonormalized)
\begin{equation}
\begin{split}
R_{\alpha,\alpha-1} 
= R_{\alpha}R_{\alpha-1}
=R_{\alpha-1}R_{\alpha},
\end{split}
\end{equation}
where
\begin{equation}
\begin{split}
R_{\beta}
=
I- 2|{q}_\beta \rangle \langle {q}_\beta |
,
\end{split}
\end{equation}
is a reflection operator over the $|{q}_\beta \rangle$ state.
Absorbing the signs of the $h_a$ into the corresponding $P_a$, let $\Lambda=\sum_{a=0}^{T-1}|h_a|$ be the LCU normalization. We can then block-encode $\Pi_{\alpha,\alpha-1}H$ using SELECT and PREPARE operators,
requiring $\lceil\log_2(2T)\rceil$ selection ancillas and producing, upon postselection, the state below with success probability $\|\Pi_{\alpha,\alpha-1}H\ket{q_\alpha}\|_2^2/\Lambda^2$ (for $\Lambda>0$):
\begin{eqnarray}
    \frac{\Pi_{\alpha,\alpha-1}H|{q}_\alpha\rangle}{\|\Pi_{\alpha,\alpha-1}H|{q}_\alpha\rangle\|_2}
    =
    |{q}_{\alpha+1}\rangle,
\end{eqnarray}
which is orthogonal to the other Lanczos states.\\

\subsubsection{Implementation challenge}

The question is to implement the composite reflection $R_{\alpha,\alpha-1} = R_\alpha\,R_{\alpha-1}$. Assume we have a state-preparation unitary $U_\alpha$ such that
\begin{equation} \begin{split}
U_\alpha\,|0\rangle = |q_\alpha\rangle.
\end{split} \end{equation}
Then, the reflection can be implemented as
\begin{equation} \begin{split}
R_\alpha
= U_\alpha\,R_{null}\,U_\alpha^\dagger
\quad
,
\quad
R_{null}=I-2|0\rangle_s\!\langle 0|_s.
\end{split} \end{equation}
The operator $R_{null}$ can be realized by a phase flip on $\lvert 0\rangle_s$ (multi-controlled $X$ and ancilla).
Combining these, the total operator would be
\begin{equation} \begin{split}
R_{\alpha,\alpha-1}
\;=\;
U_\alpha\,R_{null}\,U_\alpha^\dagger\;
U_{\alpha-1}\,R_{null}\,U_{\alpha-1}^\dagger.
\end{split} \end{equation}

The practical difficulty is that implementing $R_{\alpha}$ exactly requires a preparation circuit $U_\alpha$ for $\ket{q_\alpha}$, inheriting the recursive state-preparation cost discussed in Sec.~\ref{sec:direct_lanczos}. A possible alternative is to approximate $R_{\alpha}$ via a parametrized ansatz
\begin{equation} \begin{split}
R_\alpha(\boldsymbol{\theta}_{\alpha})
\;=\;
U_\alpha(\boldsymbol{\theta}_{\alpha})\,R_{null}\,U_\alpha(\boldsymbol{\theta}_{\alpha})^\dagger,
\end{split} \end{equation}
where $|q_\alpha(\boldsymbol{\theta}_{\alpha})\rangle
=
U_\alpha(\boldsymbol{\theta}_{\alpha})\,|0\rangle$ and the variational parameters $\boldsymbol{\theta}_\alpha$ are optimized for instance so that $q_\alpha(\boldsymbol{\theta}_\alpha)$ tends to be co-linear to $q_{\alpha}$ and orthogonal to the other states ($\langle q_{\alpha-1} \,\vert\, q_\alpha(\boldsymbol{\theta}_\alpha)\rangle $ small).
The optimized operator $R_\alpha(\boldsymbol{\theta}_\alpha)$ can then replace the exact reflection $R_\alpha$ inside the LCU block-encoding.
The same can be done for $R_{\alpha-1}$.
This approach is closely related to the variational construction discussed in Sec.~\ref{subsec:iterative_construction}.



\bibliography{references}

\end{document}